\documentclass[aps,prl,reprint]{revtex4-2}
\usepackage{graphicx}
\usepackage{color}
\usepackage{dcolumn}
\usepackage{bm}
\usepackage{comment}

\usepackage{physics}
\usepackage{siunitx}
\usepackage[version=4]{mhchem}
\usepackage[normalem]{ulem} 
\usepackage{comment}

\begin{document}

\preprint{APS/123-QED}

\author{Riccardo Foffi }
\affiliation{Department of Physics, Sapienza Universit\`a di Roma, Piazzale Aldo Moro, 2, 00185 Rome, Italy}
\author{Francesco Sciortino}
\affiliation{Department of Physics, Sapienza Universit\`a di Roma, Piazzale Aldo Moro, 2, 00185 Rome, Italy}

\date{\today}

\begin{abstract}
We numerically  investigate the metastable equilibrium structure of deep supercooled and glassy water under pressure, covering   the range of densities corresponding to the experimentally produced high-density and very-high-density  amorphous phases.  At $T=188$ K, a continuous increase in density  
is observed on varying pressure from 2.5 to 13 kbar, with no signs of first-order transitions. Exploiting a recently proposed approach to the analysis of the radial distribution function --- based on topological properties of the hydrogen-bond network --- we are able to identify well-defined local geometries that involve  pair of molecules separated by multiple hydrogen bonds, specific of the  high and very high density structures.

\end{abstract}

\title{On the structure of high-pressure supercooled and glassy water}
\maketitle

The structure of the different amorphous forms of water, their interconversions and their connection to the liquid state are fascinating topics, which keep capturing the attention of the scientific community \cite{mishima1998relationship,debenedetti2003supercooled,amann-winkel2016colloquium, handle2017supercooled, tanaka2020liquid}.
One of these amorphous states, the low-density amorphous ice (LDA), can be generated via a multiplicity of processes (vapour deposition~\cite{burton1935crystal}, hyper-quenching~\cite{mayer1985new}, heating/decompression of other disordered forms of ice~\cite{winkel2008water}) and it is now well characterised~\cite{debenedetti2003supercooled,bowron2006local,handle2018experimental}.
LDA has a density \( \rho = \SI{0.94}{\g\per\centi\m\cubed} \) at ambient pressure ($P$) and temperature  \( T=\SI{80}{\K} \), and it is characterised by locally tetrahedral structures in which each molecule accepts and donates two hydrogen bonds (HBs), the disordered equivalent of ice I~\cite{amann-winkel2016colloquium}.
The results on the high-density forms of amorphous water, for which density and structure seem to be somehow dependent on production process and annealing protocols, are much more controversial.
The most recent picture describes two variants of high-density amorphous ice (HDA), called un-annealed HDA (uHDA) and expanded HDA (eHDA), characterised by similar densities $\rho_\text{{HDA}}$,  1.15 and \SI{1.13}{\g\per\centi\m\cubed} respectively, along with a third form, the very-high-density amorphous ice (VHDA) with \( \rho_\text{VHDA} = \SI{1.26}{\g\per\centi\m\cubed} \).
uHDA is the material resulting from compression of ice I\textsubscript{h} at \SI{77}{\K} to above \si{\giga\pascal} pressures~\cite{mishima1984melting}, while eHDA can be obtained, among other methods, from the decompression of VHDA at \SI{140}{\K} to pressures lower than \( \approx \SI{0.1}{\giga\pascal} \)~\cite{winkel2008water}.
Despite their structural similarities, eHDA appears to have a significantly enhanced thermal stability with respect to uHDA~\cite{amann-winkel2013water,fuentes-landete2019nature}.
VHDA can be prepared either from ice I\textsubscript{h} by pressure-induced amorphization above \SI{1.2}{\giga\pascal} at temperatures \( 130 < T < \SI{150}{\K} \)~\cite{mishima1996relationship} or by annealing uHDA to \( T > \SI{160}{\K} \) at \si{\giga\pascal} pressures~\cite{loerting2001second}.
More details on the preparation routes of amorphous states can be found in recent reviews~\cite{loerting2011how,finney2015water,handle2017supercooled}.

Recent experimental results
 show that eHDA can be reversibly transformed into VHDA, suggesting that despite significant structural differences eHDA and VHDA might be linked to one another via a continuous thermodynamic path~\cite{handle2018experimentala}.
In other experiments a seemingly jump-like transition between HDA and VHDA was instead identified~\cite{loerting2006relation,loerting2006amorphous, salzmann2006isobaric}.

Experimental and numerical investigations have established that the most prominent structural difference between the low- and high-density forms of amorphous water is an inward collapse of the second hydration shell, a transformation that mimics the structural changes observed upon compression in the liquid state~\cite{soper2000structures,mariedahl2018xray,tanaka2019revealing,martelli2020connection}.
Less understood are the differences between HDA and VHDA; a further collapse of the second shell has been clearly observed experimentally~\cite{finney2002structure,mariedahl2018xray,amann-winkel2019structural}, but it is not clear if additional novel structural features exist that distinguish VHDA from HDA. Also, a more general question concerns whether, and to what extent, VHDA and HDA actually behave as distinct materials. 
This Letter focuses  exactly on these two points: (a) how can we interpret the structural changes and (b) are eHDA and VHDA separated by a first order transition?
We find that the simulated liquid explores configurations that accurately mimic the structural changes observed experimentally between HDA and VHDA, implying 
a deep connection  between pressure-prepared
amorphous ice and deeply supercooled water~\cite{martelli2020connection}, oppositely to  conclusions reported in other studies~\cite{shephard2017highdensity,tulk2019absence}.
We also find that at such low $T$  the equation of state does not show 
any  unstable region in this density range, consistent with the hypothesis of a continuity of states between eHDA and VHDA. Finally, we characterise the structural crossover from HDA to VHDA by identifying a connection between specific local geometries and topological  properties of the HB network, providing a quantitative characterization of what is generically (and possibly improperly) referred to as the collapse of the second neighbour shell.

A typical experimental procedure to generate HDA and VHDA requires the compression of ice I\textsubscript{h} at low \( T \), which transforms, via a mechanical instability~\cite{tse1999mechanisms}, into a disordered material that is later recovered at ambient $P$.  Repeating   this process on the computer, while often numerically  implemented~\cite{martelli2020connection, poole1992phase, giovambattista2005phase, giovambattista2005relation, martonak2005evolution, handle2019glass} requires pressure rates significantly faster than the experimental counterpart, allowing for the possibility that the resulting configurations heavily depend on the protocol.
Our numerical approach to generate glass samples exploits the picture of the glass as a quenched liquid. We investigate high-density liquid configurations of TIP4P/Ice water~\cite{abascal2005potential} along a cold ($T=\SI{188}{\K}$) isotherm, right below the model liquid-liquid critical temperature~$T_c=\SI{188.6}{\K}$~\cite{debenedetti2020second}, from 2.5 to \SI{13}{\kilo\bar}, corresponding to a density range from 1.1 to \SI{1.3}{\g\per\centi\m\cubed}. The results for 2.5 and \SI{4}{\kilo\bar} are reproduced from Ref.~\cite{foffi2021structural}. 
 Performing novel numerical simulations longer than 10--\SI{25}{\micro\s}, it is still possible to reach (metastable) equilibrium at this $T$,  eliminating any concern of history dependence and out-of-equilibrium effects.
The associated glass structures are provided by the inherent structures
(IS), the local energy minima reached via a constant-volume steepest descent procedure 
which  removes vibrational  distortions~\cite{stillinger2015energy}.
All the numerical procedures are discussed in detail in the S.M. (which includes Refs.~\cite{abraham2015gromacs,frenkel2002understanding,parrinello1981polymorphic,hess1997lincs,allen2017computer,berendsen1984molecular,kumar2007hydrogen,saito2018crucial}).

\begin{figure}
	\centering
	\includegraphics[width=8.6cm]{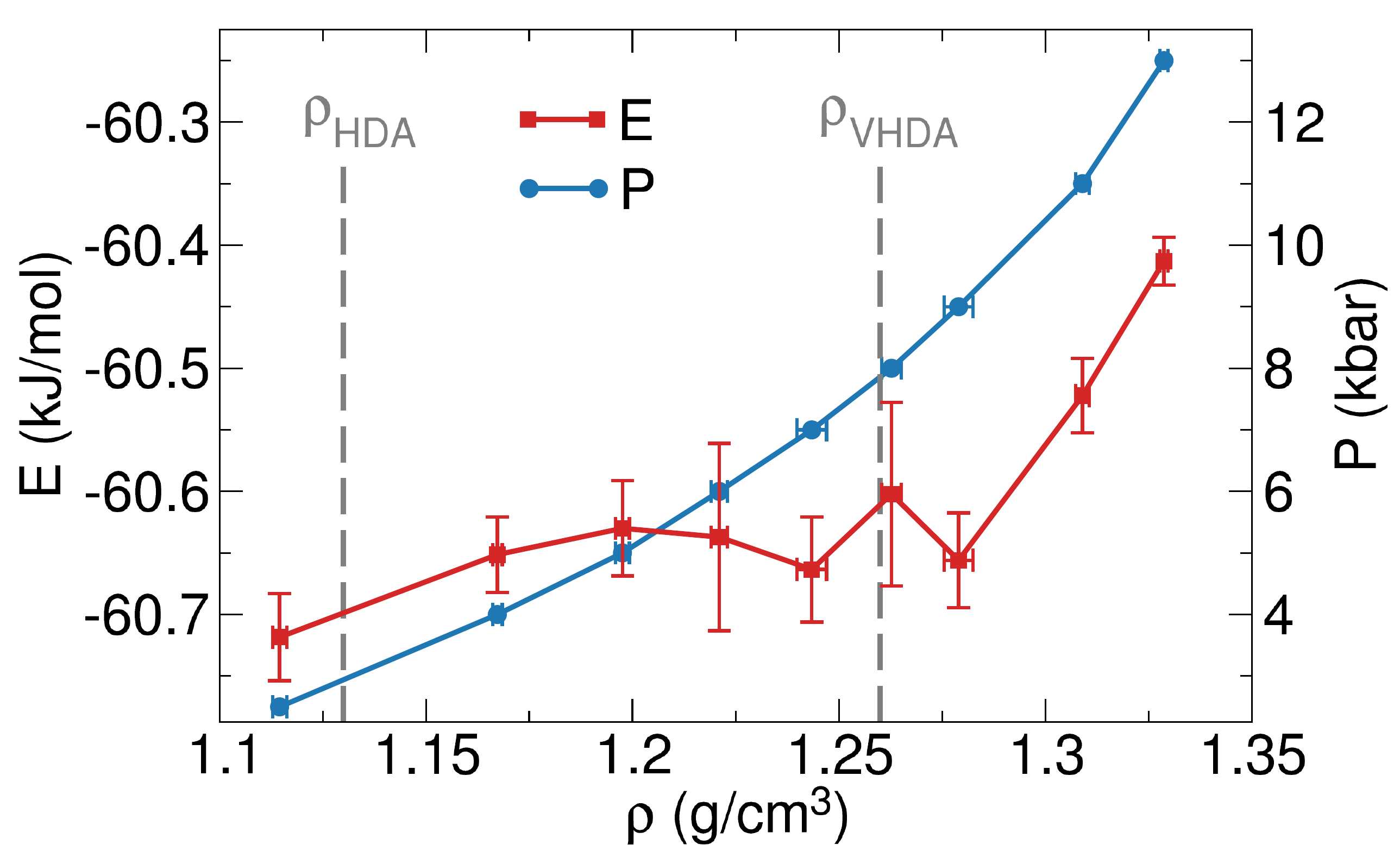}
	\caption{Thermodynamics of TIP4P/Ice water from MD simulations along the \SI{188}{\K} isotherm.
	 Equation of state (right $y$-axis) and potential energy (per molecule, left $y$-axis) as a function of density. Grey dashed lines show the eHDA and VHDA experimental $\rho$ at \SI{80}{\K} and \SI{1}{\bar} for reference. Error bars are estimated as the variance of the averages over 10 distinct time intervals in the production run.}
	\label{fig:thermo}
\end{figure}

Figure~\ref{fig:thermo} shows the potential energy (per molecule) \( E \) as a function of density (\( \rho \)) and the equation of state \( P \)  vs. \( \rho \) at \( T = \SI{188}{\K} \).
In the explored  range, \( P(\rho) \)  has positive concavity and no discontinuities, establishing the absence of any thermodynamic transition between two distinct dense liquid forms along this isotherm.  
Quite interesting, however, is the density dependence of the potential energy $E$, which shows a  region between 1.12 and \SI{1.25}{\g\per\centi\m\cubed}  (the typical density values of eHDA and VHDA) in which $E$ weakly depends on $\rho$, suggesting that the structural changes taking place on increasing pressure have a small energetic cost.  Only beyond $\rho_\text{VHDA}$, $E$ grows significantly with $\rho$. 

\begin{figure}
	\centering
	\includegraphics[width=8.6cm]{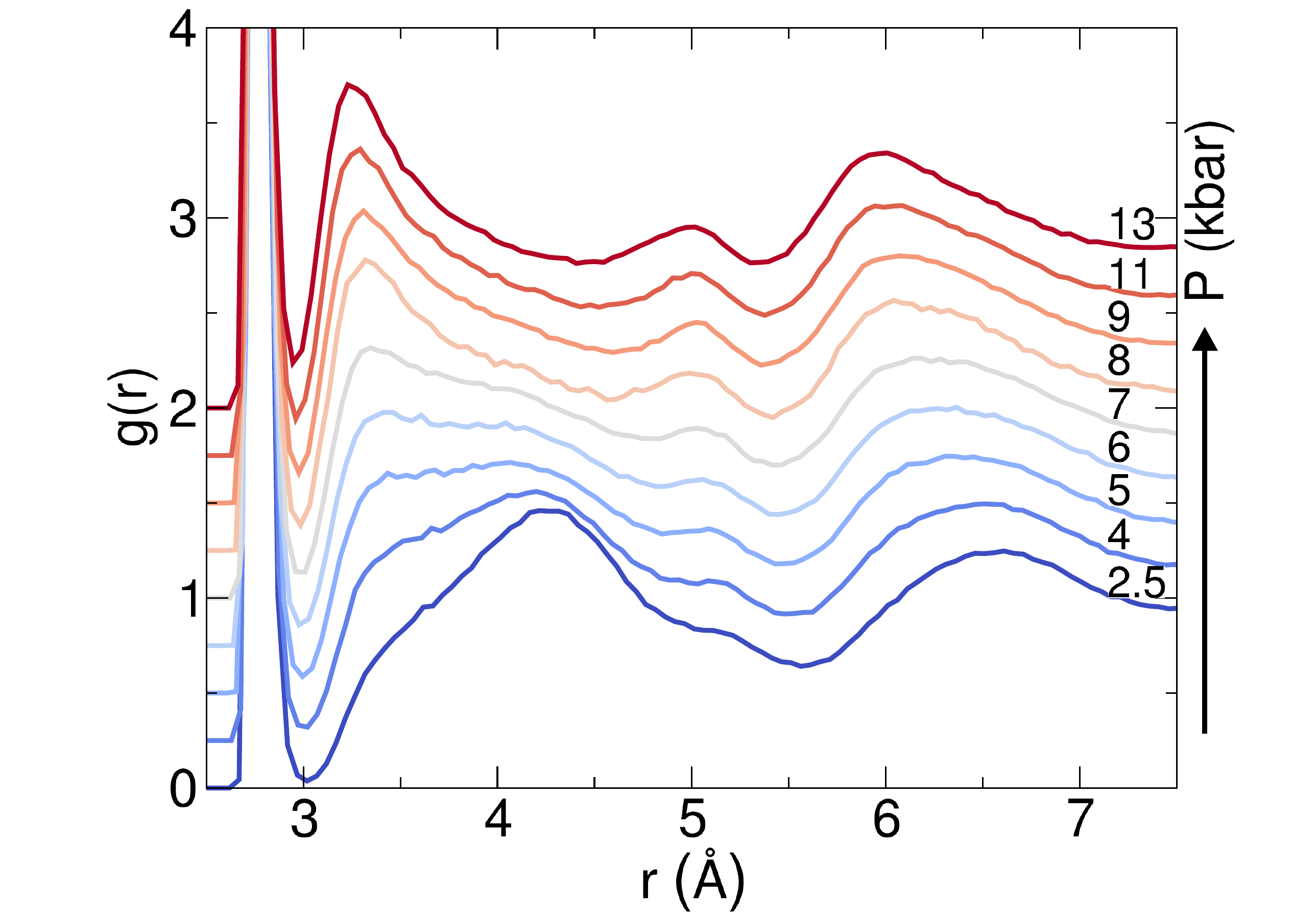}
	\caption{Oxygen-oxygen radial distribution function  evaluated in the 
	IS along the \SI{188}{\K} isotherm from 2.5 to \SI{13}{\kilo\bar}. The 
	evaluation of the spatial correlations in the IS highlights the 
	evolution of the structural features.
	Successive \( g(r) \) curves are shifted by 0.25 on the vertical axis}
	\label{fig:structureIS}
\end{figure}

Fig.~\ref{fig:structureIS} shows the continuous smooth evolution of the calculated oxygen-oxygen radial distribution function \( g(r) \). 
On increasing \( P \), the concentration of interstitial molecules
(\( r\approx3.2-\SI{3.5}{\angstrom} \)) increases significantly, while the tetrahedral peak (\( r\approx\SI{4.4}{\angstrom} \)) is suppressed.
Two additional features characterise the  evolution: the development of a novel peak at \( r\approx\SI{5}{\angstrom} \) and the intensity crossover from \( r\approx\SI{6.7}{\angstrom} \) to \( \approx\SI{6}{\angstrom} \). 
The corresponding evolution of the structure factor~\cite{hansen2013theory} is reported in the S.M.

\begin{figure}
	\centering
	\includegraphics[width=8.6cm]{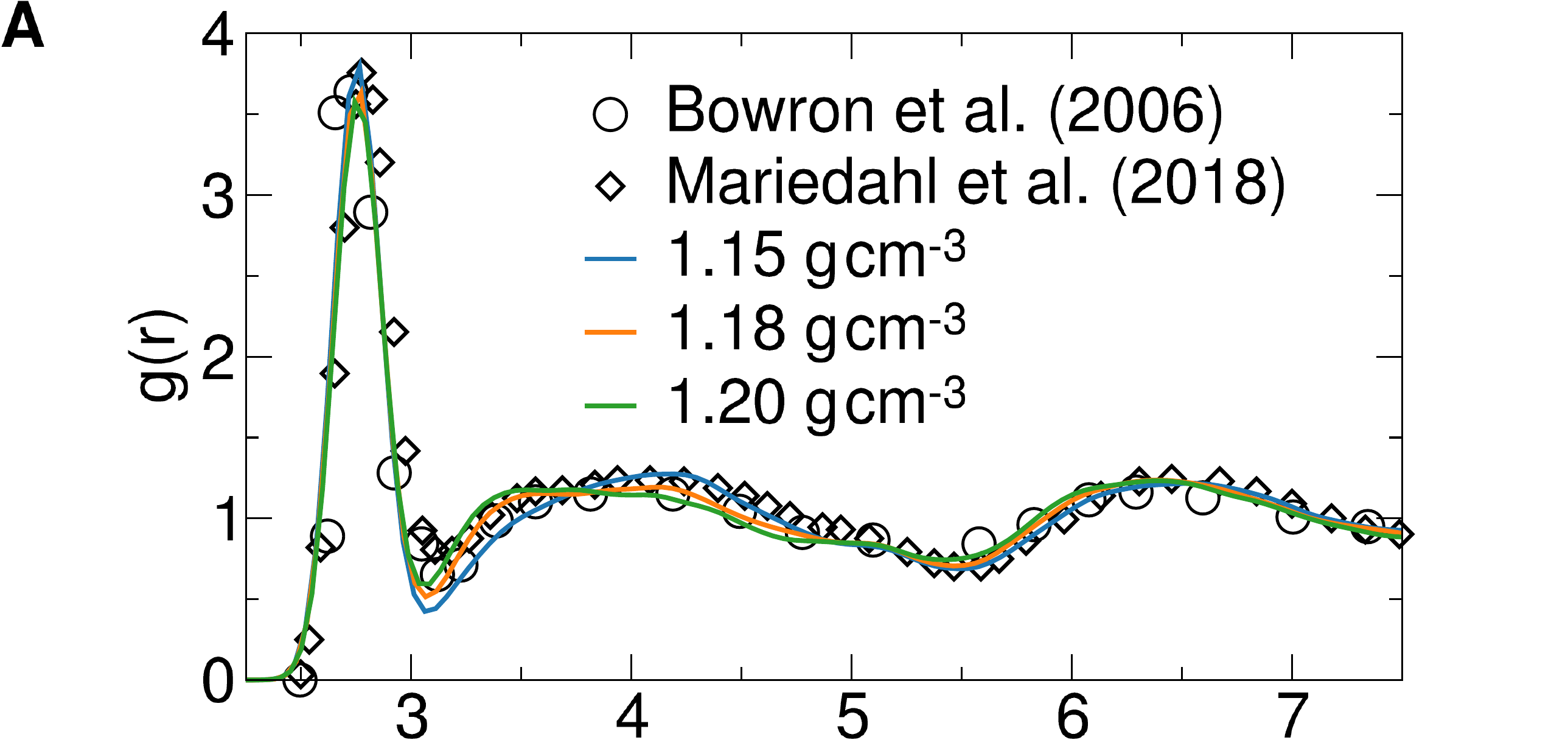}
	\includegraphics[width=8.6cm]{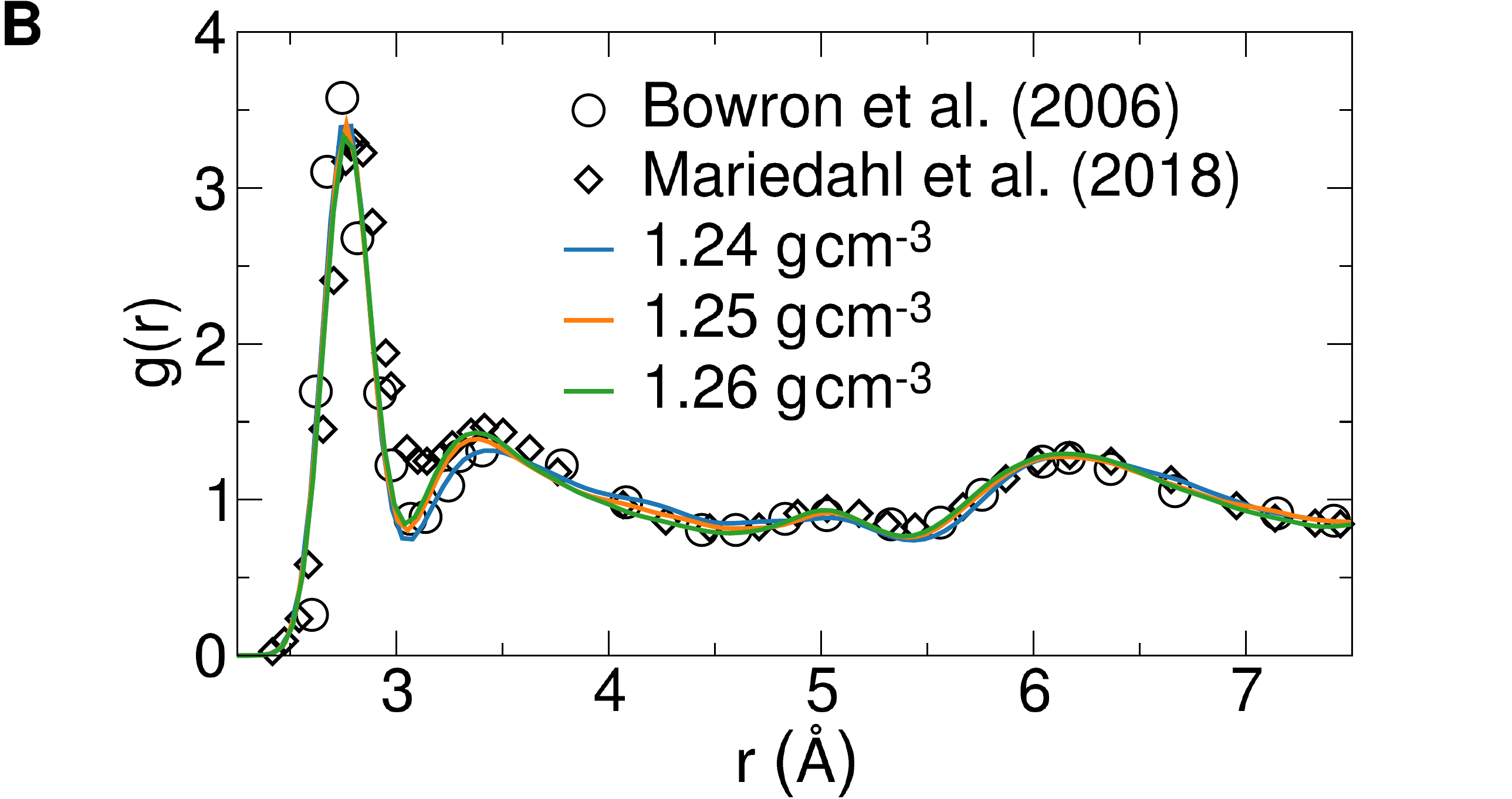}
	\caption{Comparison between experimental (symbols) and numerical (lines) $g(r)$ in (a) eHDA and (b) VHDA at \( T=\SI{80}{\K} \) and \( P=\SI{1}{\bar} \). 
	The densities indicated in the label refer to the values at ambient pressure.
	}
	\label{fig:rdf-exp}
\end{figure}

Before interpreting the structural features,  we validate the  numerical  
results  
 comparing the structure of the glasses generated by quenching equilibrium liquid configurations with  available experimental data of 
eHDA and VHDA structures from \citet{mariedahl2018xray} and 
\citet{bowron2006local}. To do so, we  quench 
configurations equilibrated at  $T=\SI{188}{\K}$ and different $P$ to \( T=\SI{80}{\K} \) and ambient pressure 
--- the same  $T$ and $P$  in Refs~\cite{mariedahl2018xray,bowron2006local}.  
Following a short MD simulation, the density and  the vibrational and rotational degrees of freedom adjust  to $T=\SI{188}{\K}$   and $P=1$ bar.  More details can be found in the S.M.
Since classical calculations neglect the quantum delocalization of the 
atoms~\cite{herrero2011pathintegral,herrero2012highdensity}, the height of the first peak of $g(r)$ is usually overestimated. 
To include in an effective way the quantum broadening of \ce{O-O} distances 
(as well as any possible broadening due to the experimental procedures)
we have  convoluted the numerical $g(r)$ with a gaussian of variance 
$\sigma_B^2=$\SI{7.1e-3}{\AA\squared}, a value obtained by matching the heights of the first peak in the numerical \( g(r) \) to the experimental values. This value is consistent with the broadening observed between path-integral and classical simulations  of the same model~\cite{herrero2012highdensity}. Note that the implemented broadening becomes
irrelevant when $r  \gg \sigma_B$ (Fig.~S8 reports the comparison without the correction). 
We compare the experimental data (which display some differences, possibly  due to different preparation histories) to numerical  $g(r)$s with $\rho$s
comparable to experimental values typical of HDA and VHDA. 
Fig.~\ref{fig:rdf-exp} shows 
that both eHDA and VHDA radial features are 
rather faithfully reproduced by TIP4P/Ice. 
The quality of the comparison enforces our confidence in the TIP4P/Ice  model to properly describe the local geometries responsible for the
structural signatures observed experimentally, as detailed in the following analysis.

To gain a better insight in the structural changes on going from eHDA to VHDA we adopt a methodology recently proposed to investigate the liquid-liquid phase transition~\cite{foffi2021structural}. We separate the contributions to the \( g(r) \) originating from pairs of molecules with  ``chemical distance'' $D$, where $D$ is defined as the minimum number of HBs  connecting the two molecules along the HB network.
An analysis based on the HB network topology depends crucially on the ability to properly identify HBs. In TIP4P/Ice, HBs can be accurately identified at all pressures (Figs.~S9--S11) adopting the Luzar-Chandler definition~\cite{luzar1996hydrogenbond}.  Furthermore, at all explored densities, more than 99.5\% of the H atoms are involved in HBs, indicating that practically all molecules donate two hydrogens for bonding (Fig.~S12).

\begin{figure}
	\centering
	\includegraphics[width=8.6cm]{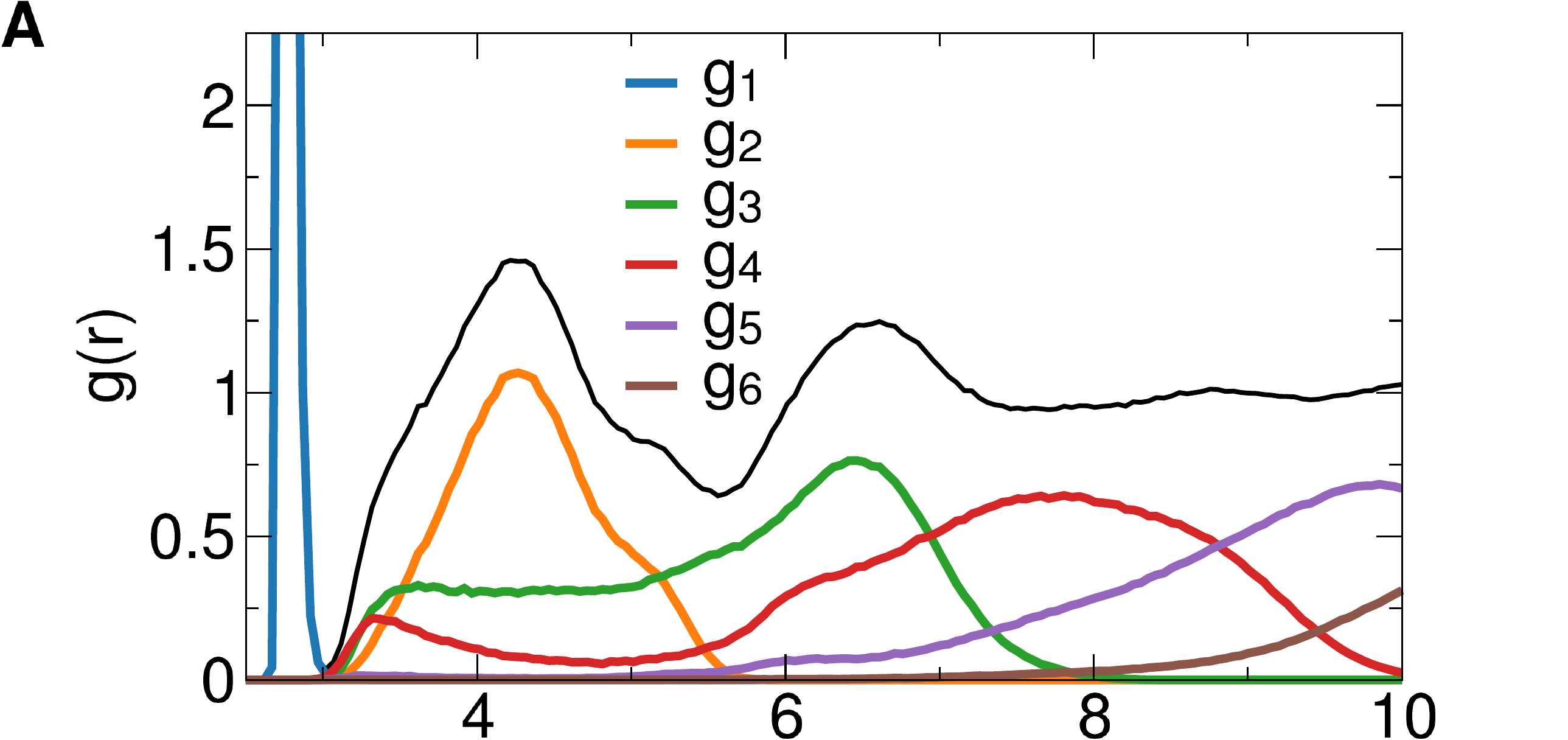}
	\includegraphics[width=8.6cm]{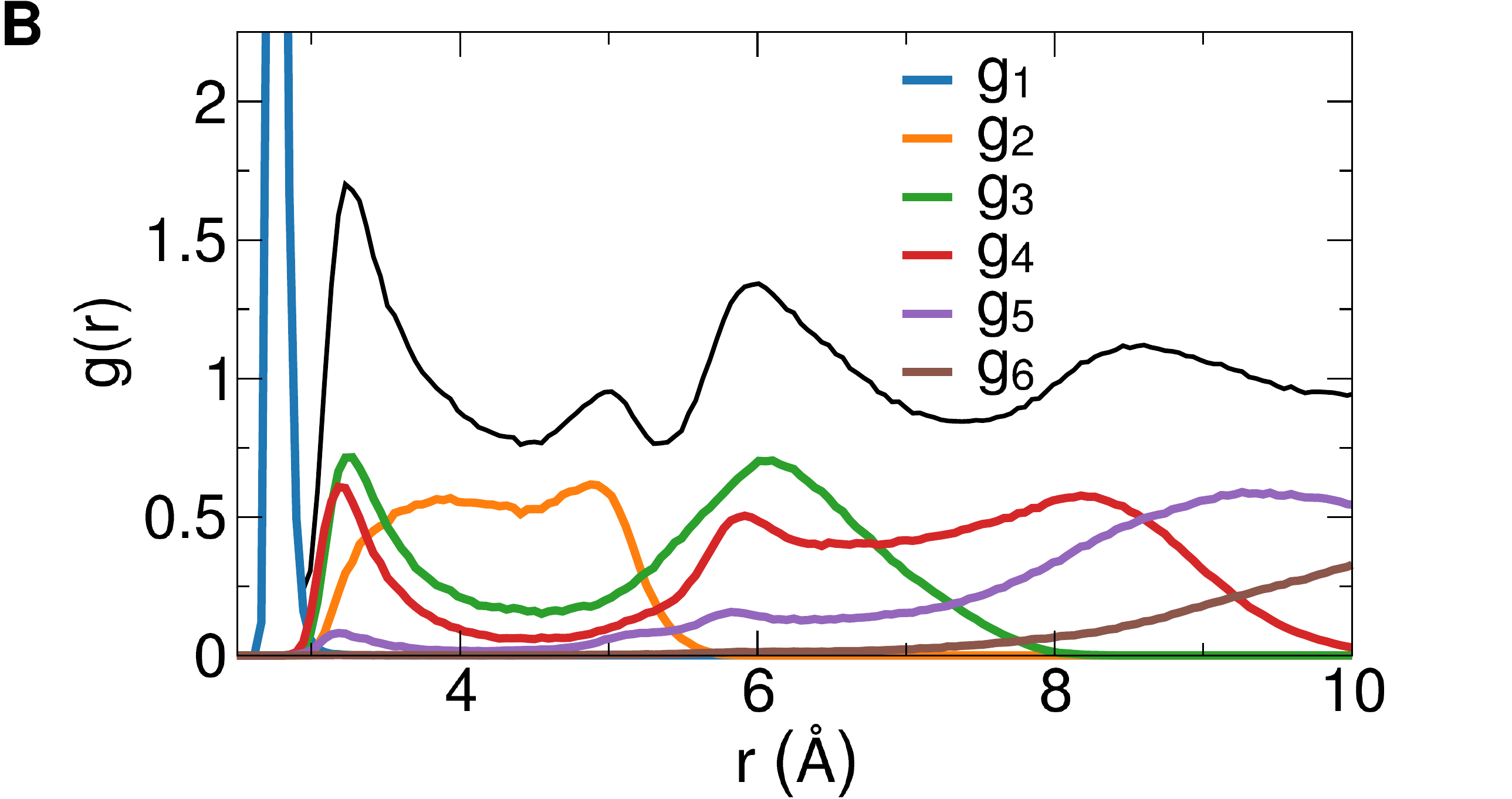}
	\caption{Radial distribution function of water (evaluated in the IS) at \SI{188}{\K} and  $P$ (a) \SI{2.5}{\kilo\bar}  and (b) \SI{13}{\kilo\bar} separated in 
	their \( g_D \) contributions $(1 \leq D \leq 6)$  from pairs of molecules at  chemical distance $D$.  
	The  black line is the sum over all $g_D(r)$, equal to the total $g(r)$  (Eq.~\ref{eq:grd}).
	}
	\label{fig:grchemdist}
\end{figure}

\begin{figure*}
	\centering
	\includegraphics[width=8.6cm]{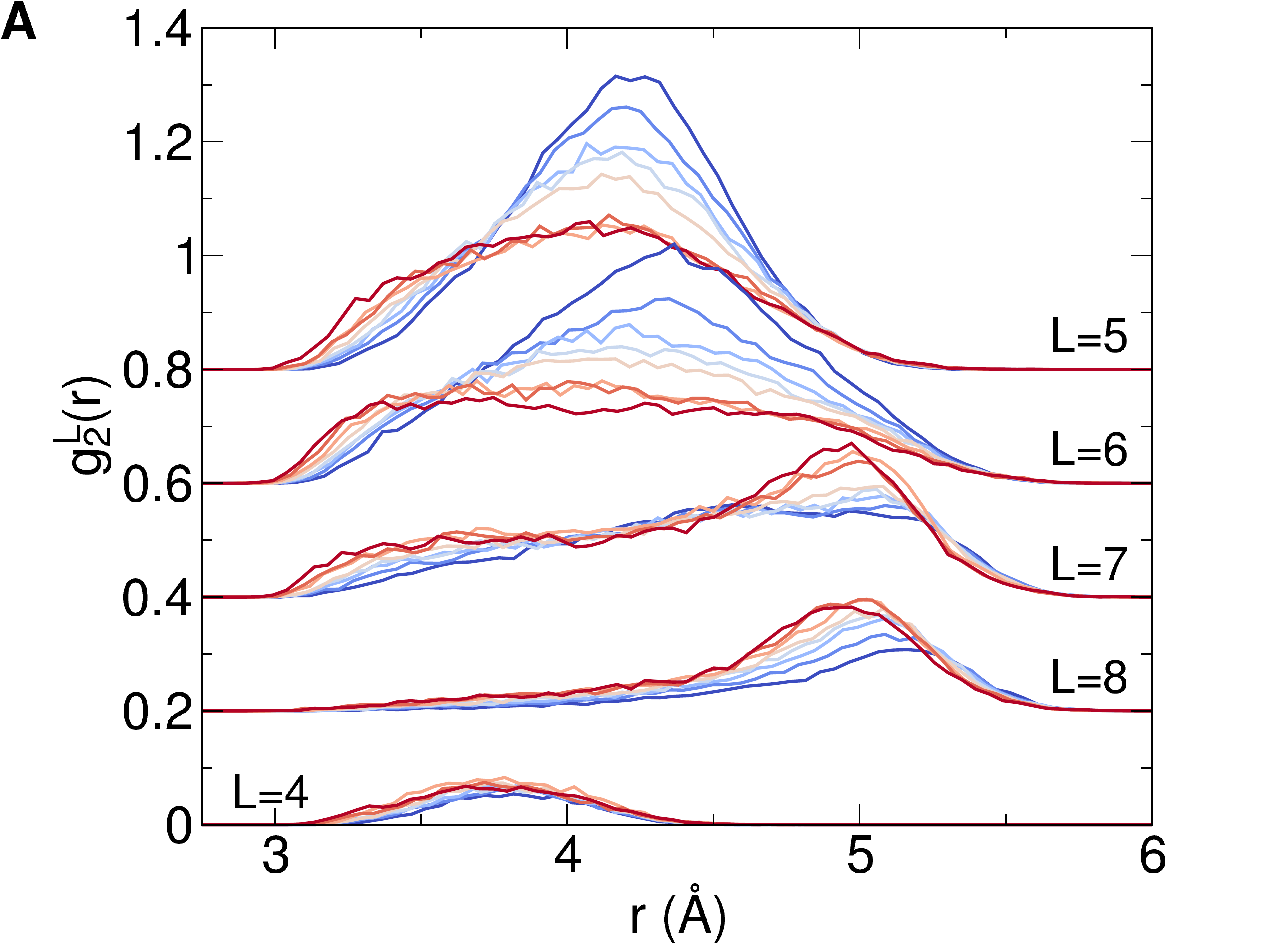}
	\includegraphics[width=8.6cm]{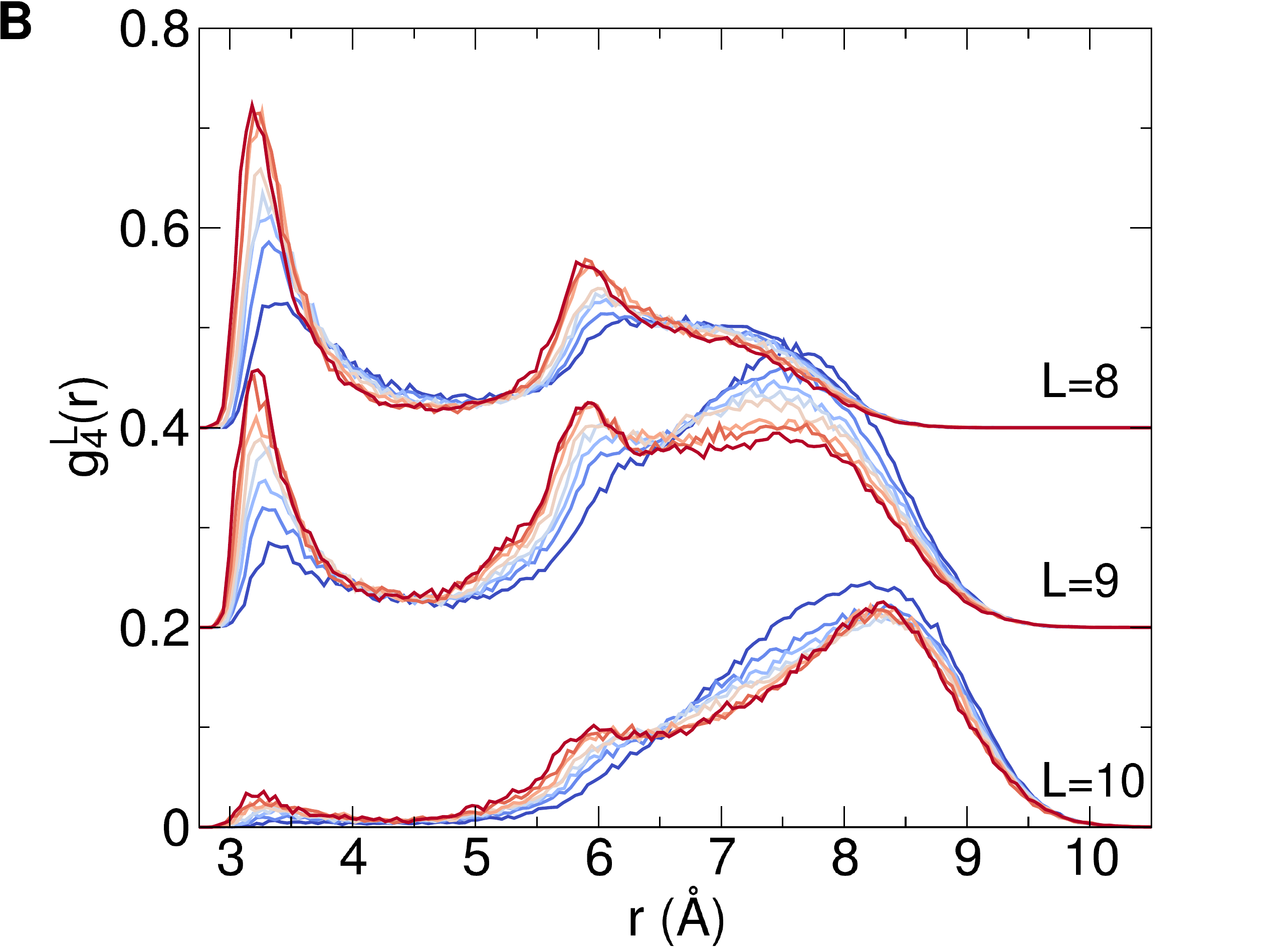}
	\caption{Contribution to the radial distribution function from  pairs 
	of molecules at (a) D=2 and (b) D=4,
	separated according to the ring 
	length $L$, along the \( T = \SI{188}{\K} \) isotherm.  Colours from 
	red to blue indicate increasing  pressure from 2.5 to
	\SI{13}{\kilo\bar}.
	Note that for $D=2$ and $L=7$ and $8$, a clear peak progressively grows at
	$r \approx \SI{5}{\angstrom}$. 
	Also note that for $D=4$, beside the growth at the interstitial 
	distance $r \approx \SI{3.2}{\angstrom}$ a further peak appears at
	$r \approx \SI{6}{\angstrom}$, most clearly for $L=8$ and $L=9$.
	}
 	\label{fig:d24l}
\end{figure*}

We connect structure and network topology  by writing
\begin{equation}
	g(r) = \sum_D g_D(r)
	\label{eq:grd}
\end{equation}
where \( g_D(r) \) is the radial distribution function evaluated among only pairs of molecules at chemical distance \( D \).
The effects of this decomposition  are shown for the lowest and highest studied $P$ (2.5 and \SI{13}{\kilo\bar}) in Fig.~\ref{fig:grchemdist}.
The partitioning reveals precise relationships between spatial and topological features, providing insight into the structural evolution that occurs upon compression and thus facilitating a classification of molecule pairs into distinct groups.
First, we observe well defined peaks at \( r\leq\SI{3.5}{\angstrom} \)  for \( D=3,4 \) and 5, clarifying that interstitial molecules, despite their close real-space distance, reaching  down to \( r\approx\SI{3}{\angstrom} \), are separated by three or more HBs.  The growth in the number of interstitial molecules,
(a phenomenon which characterizes the transition from the low to the high density liquid~\cite{foffi2021structural}) is significantly enhanced on going from HDA-like to VHDA-like densities.
Figure~\ref{fig:grchemdist} also shows that  the peak at \( r\approx\SI{5}{\angstrom} \)  can be associated to  molecules separated only by two HBs (\( D=2 \)) and that the peak at \( r\approx\SI{6}{\angstrom} \) originates from pairs  with chemical distance \( D=4 \).
Indeed, while \( D=3 \) provides a major contribution to 
$g(r\approx\SI{6}{\angstrom})$, only in \( D=4 \) a new peak arises, continuously growing upon compression from high to very-high densities.

Finally, $g_2(r)$ shows a complete disappearance of the tetrahedral  peak at \SI{4.4}{\angstrom} --- the landmark of the LDL phase, still quite intense in the HDL at coexistence \cite{foffi2021structural} --- which progressively transforms into a broad  featureless distribution, indicating a significant broadening of the \( \ce{O}\ce{\hat{O}}\ce{O} \) angle of H-bonded triplets of molecules.

To pin down the local geometries responsible for the peaks, 
$g_D(r)$ can be further decomposed into contributions from ``rings'' of different length~\cite{foffi2021structural}.
To each pair of molecules contributing to $g(r)$ we now associate, in addition to the spatial distance \( r \) and chemical distance \( D \), also a ring length $L$.
The ring is selected by joining the two shortest non-intersecting HB-paths connecting the selected pair (see also Fig.~S15 for an example). Then each $g_D$ can be written as
\begin{equation}\label{eq:gdl}
	g_D(r)=\sum_{L} g_D^L(r)
\end{equation}
where $g_D^L$ indicates the contribution to $g(r)$ from all pairs of molecules with chemical distance $D$ and ring length $L$.

\begin{figure}
	\includegraphics[width=3.8cm]{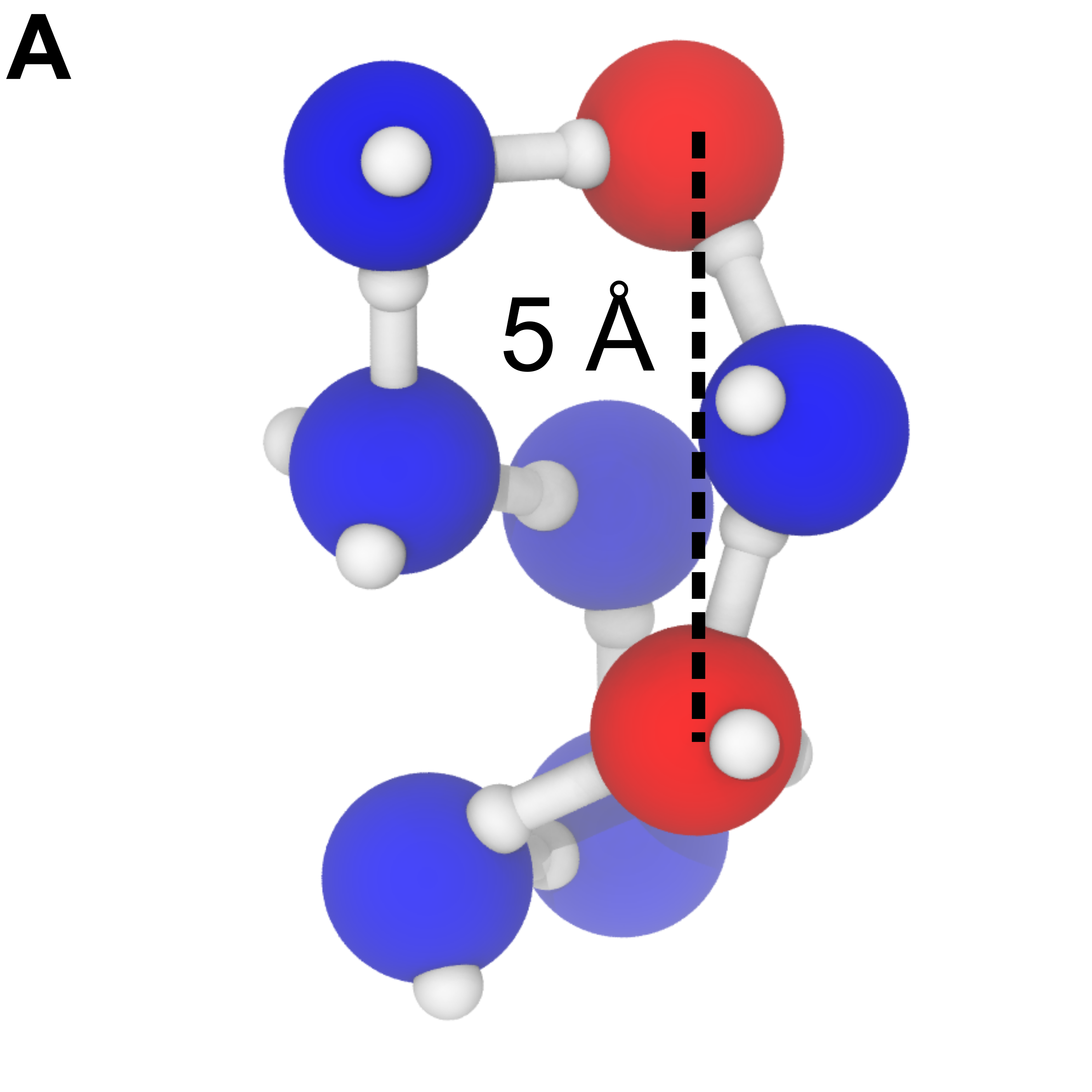}
	\includegraphics[width=3.8cm]{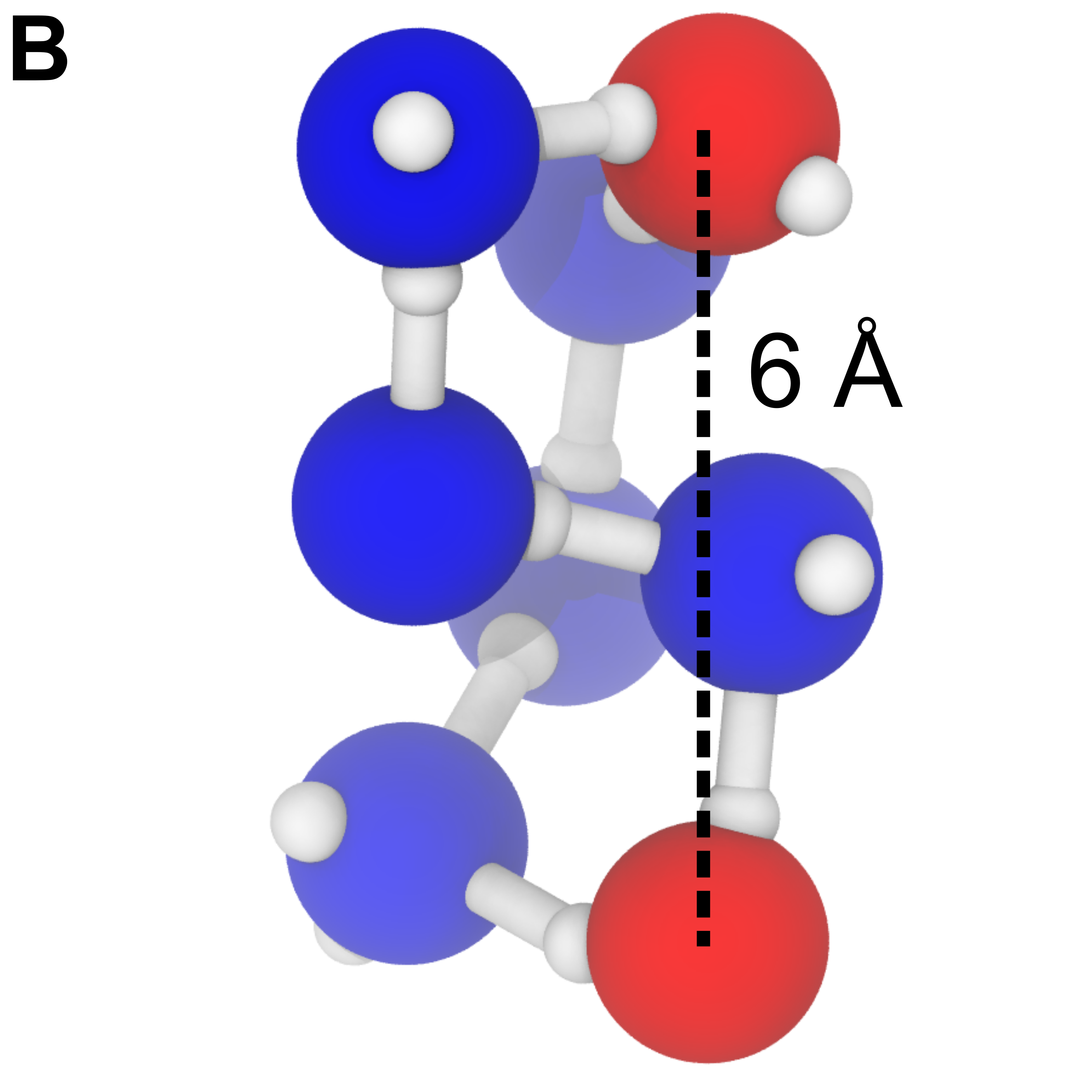}
	\includegraphics[width=3.8cm]{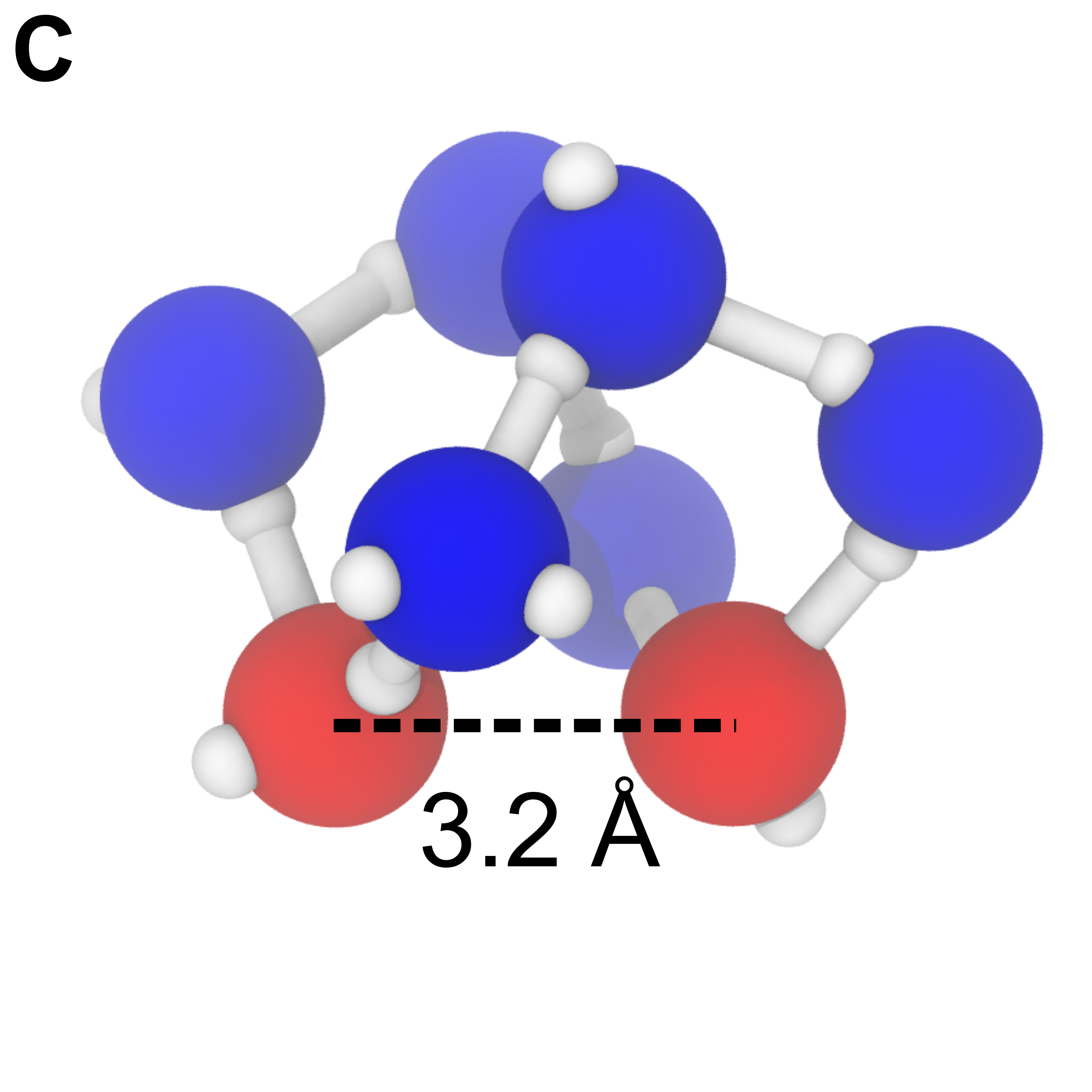}
	\caption{Cartoon representation of the three local geometries that 
	characterise water under high pressure: (a) pairs at $D=2$ and $L=8$  
	(D2L8) and (b,c)  pairs with $D=4$ and $L=8$ (D4L8).
	D2L8 pairs are separated by a single molecule with a wide angle between 
	the two HBs, and contribute to the \( g(r) \) peak at
	\( r\approx\SI{5}{\angstrom} \).
	D4L8 pairs contribute to the peaks at
	\( r\approx\SI{6}{\angstrom} \) and \( r\approx\SI{3.2}{\angstrom} \).
	In the first case (b), the pair is almost-collinear with a third	
	molecule H-bonded to only one of them;
	in the latter (c) the folding back of the ring allows the pair to get
	very close despite the large chemical distance.}
	\label{fig:rings}
\end{figure}

This analysis applied to $g_2(r)$, detailed in Fig.~\ref{fig:d24l}(a), reveals that the peak at \( r\approx\SI{5} {\angstrom} \) originates from pairs at \( D=2 \) which are part of rings of length \( L = 7 \) and $8$. An example of the corresponding molecular arrangement is shown in Fig.~\ref{fig:rings}(a). Since the length of a HB is strongly constrained around \( \approx\SI{2.8}{\angstrom} \), these types of configurations at \( D=2 \) are associated to typical \( \ce{O}\ce{\hat{O}}\ce{O} \) angles of about \SI{135}{\degree}. Such large angles (compared to the tetrahedral one) propagate in the HB network, promoting  the formation of  \( L \geq 7 \) rings.  
The same strategy (Eq.~\ref{eq:gdl}) also allows us to associate the peak at \( \approx\SI{6}{\angstrom} \) to pairs at \( D=4 \) for different values of \( L \); the decomposition of \( g_4(r) \) is reported in Fig.~\ref{fig:d24l}(b).
Pairs at \( r\approx\SI{6}{\angstrom} \)
%
%
are found in rings with a square-like segment such that they are almost-collinear with a third molecule, H-bonded to only one of them, as seen in Fig.~\ref{fig:rings}(b). 
Pairs with $D=4$  and $L=8$  also provide a major contribution to the interstitial region with a crystal-clear peak 
shifting from 3.5 to \SI{3.2}{\angstrom} with increasing \( P \), signaling a drastic network restructuring linked to the interpenetration of bond-coordination shells.  An example of this local geometry is provided in Fig.~\ref{fig:rings}(c).
Movies~S1--S3 included in the S.M. facilitate the visualization of these geometric arrangements.

In summary, by exploring with lengthy simulations the 
high-density liquid in TIP4P/Ice at $T=\SI{188}{\K}$ (below the model liquid-liquid $T_c$), we have demonstrated that the liquid structure continuously evolves from eHDA to VHDA, supporting an interpretation of  water high-density glasses based on thermodynamic continuity.
Both eHDA and VHDA can be seen as the kinetically arrested counterparts of a single high-density liquid, which is mirrored by eHDA close to the liquid-liquid transition pressure, and progressively more by VHDA as $P$ is increased.
Together with the experimentally established connection between the low density liquid and LDA~\cite{kringle2020reversible,kringle2021structural}, our finding reinforces the hypothesis, based on the liquid-liquid critical point  idea~\cite{poole1992phase,palmer2018advances,kim2020experimental}, that LDA and HDA-VHDA  are the amorphous phases associated to the low and high density liquids, with no need to invoke a derailed crystallization pathway~\cite{shephard2017highdensity,tulk2019absence}.
 
We have also identified the geometrical origin of the
structural changes taking place on going from eHDA to VHDA: (i) interstitial molecules arising from long rings
with a ``folded'' structure,
bringing molecules separated by three or four HBs close by in space; (ii)  progressive distortion of the  \( \ce{O}\ce{\hat{O}}\ce{O} \) angle,  favouring larger ring sizes (and responsible for the
peak at $r \approx \SI{5}{\angstrom}$); (iii) specific structural motifs  in rings of length \( L \geq 8 \)
such that molecules separated by four HBs
remain at distance \( \approx\SI{6}{\angstrom} \).

\begin{acknowledgments}
We acknowledge discussions with Prof. Livia Bove  and John Russo and support from MIUR PRIN 2017 (Project 2017Z55KCW).
We also thank HPC-CINECA for providing  computational resources.
\end{acknowledgments}

%

\end{document}



\author{Riccardo Foffi }
\affiliation{Department of Physics, Sapienza Universit\`a di Roma, Piazzale Aldo Moro, 5, 00185 Rome, Italy}
\author{Francesco Sciortino}
\affiliation{Department of Physics, Sapienza Universit\`a di Roma, Piazzale Aldo Moro, 5, 00185 Rome, Italy}


\title{Supplementary Materials for: On the structure of high-pressure supercooled and glassy water}
\maketitle

\section{Description of the simulations}
This study is based on a set of classical molecular dynamics (MD) simulations of \( N=1000 \) water molecules interacting via the TIP4P/Ice potential~\cite{abascal2005potential} in the NPT ensemble, using GROMACS 5.1.4~\cite{abraham2015gromacs} in single precision.
A cubic simulation box with periodic boundary conditions was adopted.
We chose a leap-frog integrator with a timestep of \SI{2}{\femto\s}.
Temperature coupling was controlled via a Nos\'e-Hoover thermostat~\cite{frenkel2002understanding} with a characteristic time of \SI{8}{\pico\s}, and pressure coupling was controlled via an isotropic Parrinello-Rahman barostat~\cite{parrinello1981polymorphic} with characteristic time \SI{18}{\pico\s}.
Molecular constraints were implemented via the LINCS algorithm~\cite{hess1997lincs} at the sixth order, and van der Waals forces have been evaluated with a cutoff distance of
\SI{0.9}{\nano\m}.
To account for electrostatic interactions  the particle-mesh Ewald method at the fourth order, with real-space cutoff \SI{0.9}{\nano\m}~\cite{allen2017computer} was selected.

\begin{figure}
	\centering
	\includegraphics[width=11cm]{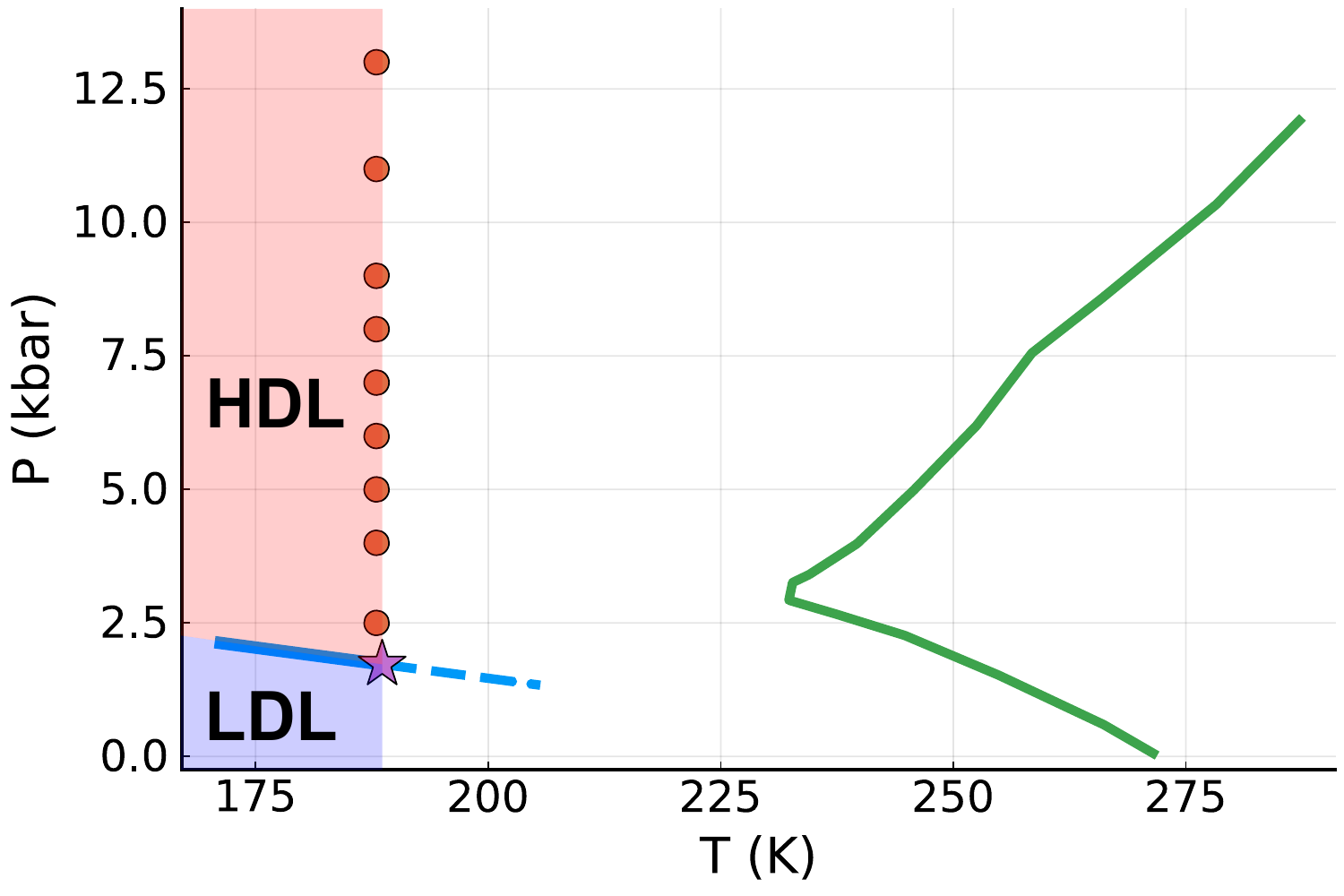}
	\caption{$P$--$T$ projection of the phase diagram of TIP4P/Ice, showing
	the explored state points (orange circles), the liquid-liquid critical
	point location estimated in Ref.~\cite{debenedetti2020second}
	(violet star), the liquid-liquid transition line from 
	Ref.~\cite{foffi2021structural} (solid blue line)
	and its extension 
	in the supercritical regime (dashed blue line) and the solid-liquid
	coexistence curve from Ref.~\cite{abascal2005potential}
	(solid green line).}
	\label{fig:phasediagram}
\end{figure}

\begin{figure}
	\centering
	\includegraphics[width=11cm]{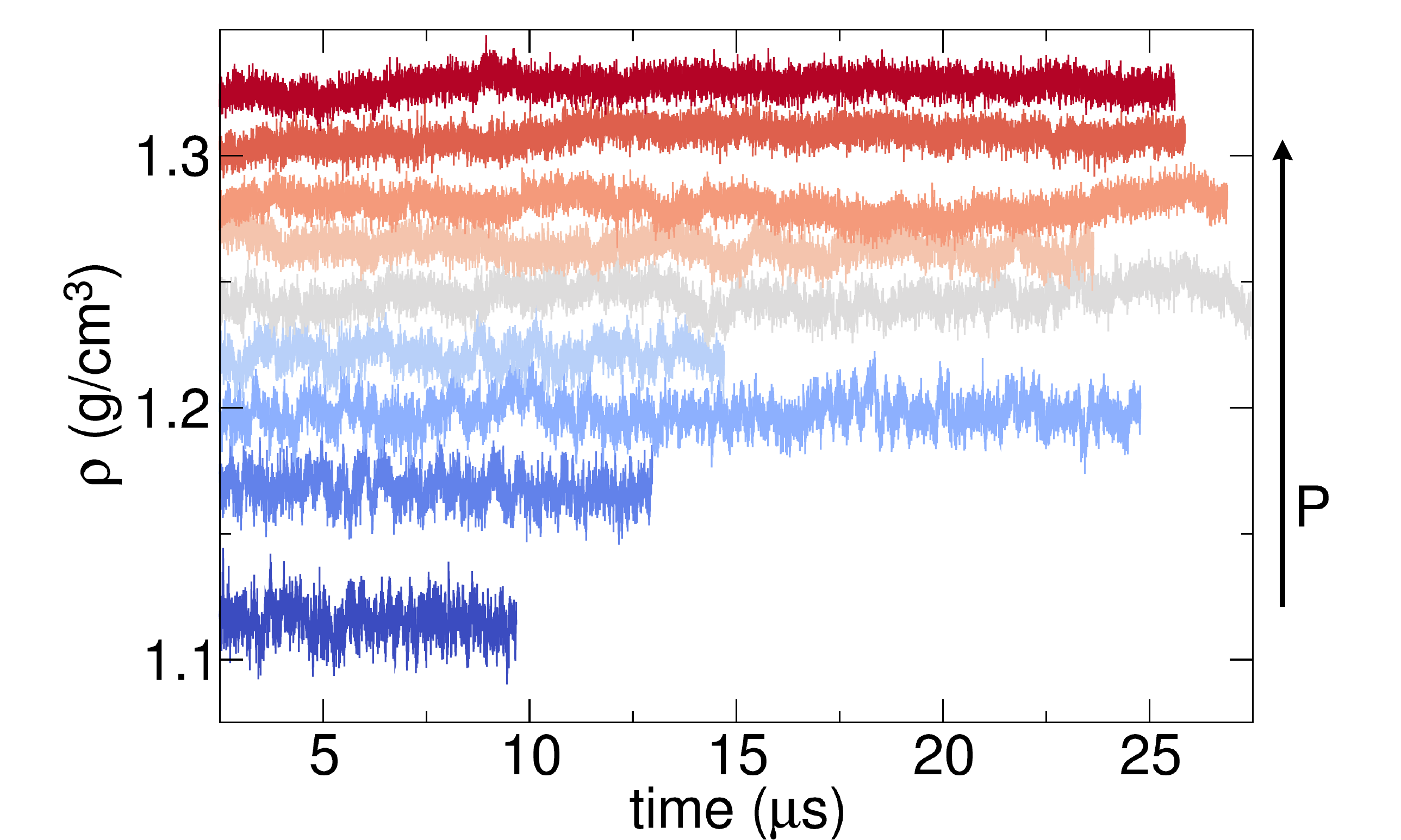}
	\caption{Evolution of density in time along the \SI{188}{\K} isotherm.
	Colors from blue to red indicate increasing pressure from 2.5 to
	\SI{13}{\kilo\bar}. The precise pressure values are listed in Table I.}
	\label{fig:density}
	\vskip 18pt
	\includegraphics[width=11cm]{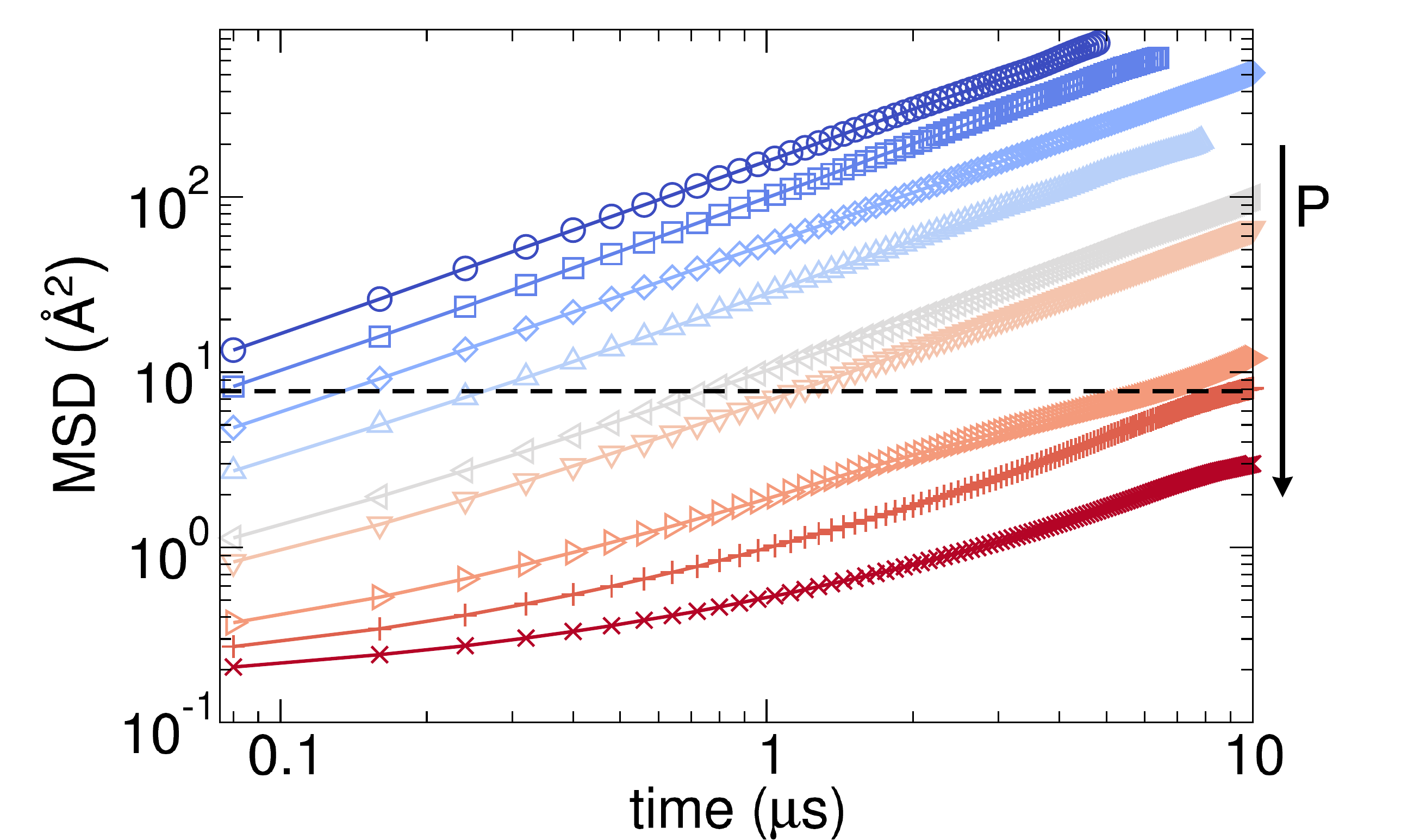}
	\caption{Mean-square displacement of water molecules during the
	production phase of the MD simulations along the \SI{188}{\K} isotherm.
	Colors from blue to red indicate increasing pressure from 2.5 to
	\SI{13}{\kilo\bar}. The precise pressure values are listed in Table I.
	The black dashed line at \SI{7.8}{\AA\squared} represents the nearest-neighbour distance squared.}
	\label{fig:msd}
\end{figure}

In Fig.~\ref{fig:phasediagram} we show the location of the state points explored in this work, with respect to other relevant features of the
phase diagram of the TIP4P/Ice model.
Equilibration and production times for these state points are reported in Table~\ref{tab:simulations}.
The simulations at pressures 2.5 and \SI{4}{\kilo\bar} have been reproduced from Ref.~\cite{foffi2021structural}, but the latter has been significantly extended.
The evolution of density with time throughout the whole simulation is shown in Fig.~\ref{fig:density}; also, the molecular mean square displacements during the production phase are reported in Fig.~\ref{fig:msd}.
Apart from the highest density (associated to pressure \SI{13}{\kilo\bar}),  molecules have diffused, on average, a distance larger than the nearest neighbour distance (the hydrogen bond length, $r\approx\SI{2.8}{\angstrom}$) during the production phase.

\begin{table}
	\centering
	\caption{Equilibration and production times of the MD simulations.}
	\label{tab:simulations}
	\begin{tabularx}{13cm}{>{\centering\quad}X>{\centering\quad}XX<{\centering\quad}}
		\toprule
		pressure / \si{\kilo\bar} & equilibration time / \si{\micro\s} & production time / \si{\micro\s} \\
		\midrule
		2.5 & 5.0 & 5.0\\
		4.0 & 6.0 & 6.5\\
		5.0 & 13.0 & 12.0\\
		6.0 & 7.0 & 8.0\\
		7.0 & 15.5 & 12.0\\
		8.0 & 13.0 & 10.0\\
		9.0 & 15.0 & 12.0\\
		11.0 & 14.0 & 12.0\\
		13.0 & 14.0 & 12.0\\
		\bottomrule
	\end{tabularx}
\end{table}

\section{Inherent structures}
\begin{figure}
	\centering
	\includegraphics[width=11cm]{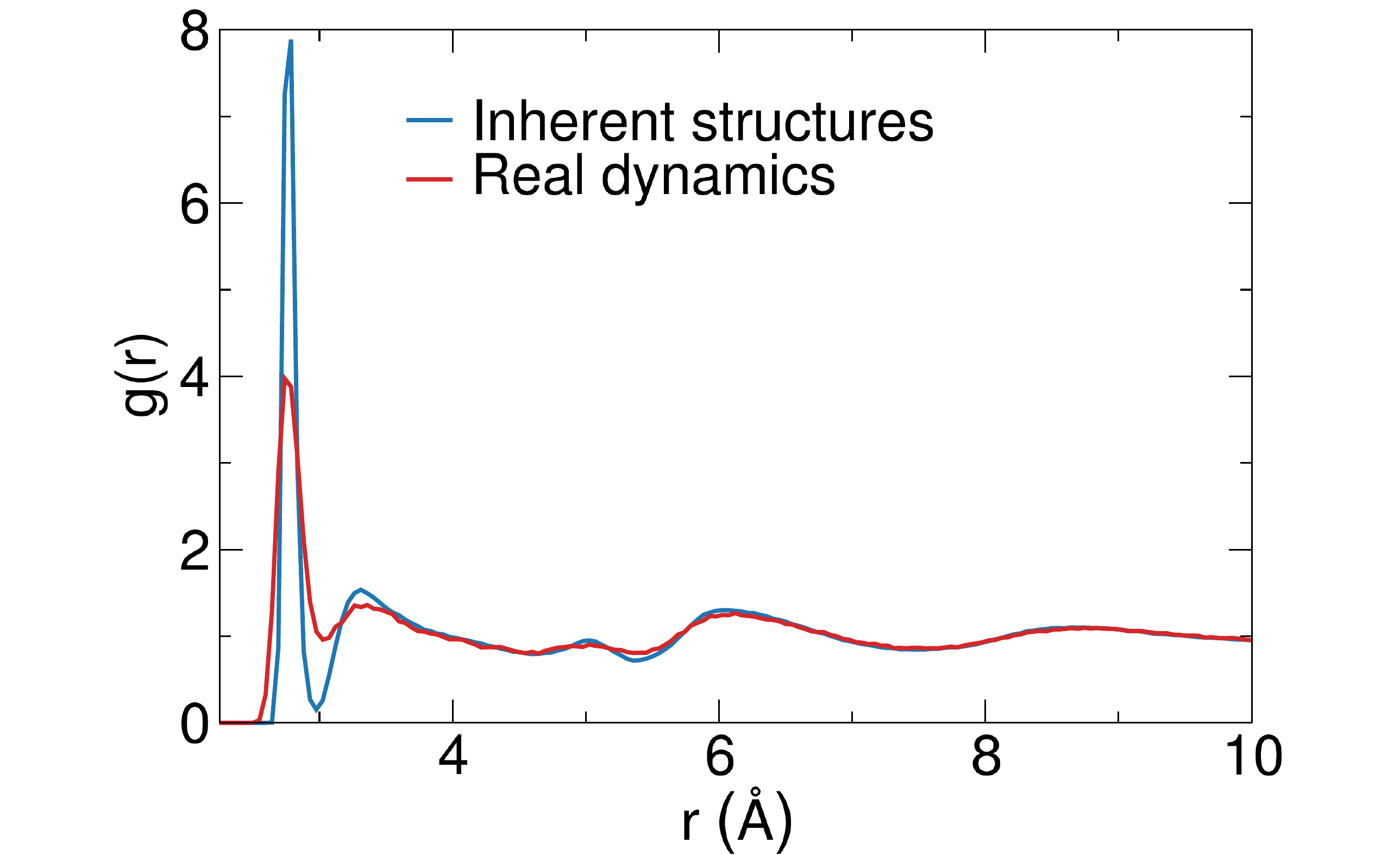}
	\caption{Comparison of the radial distribution function of the liquid
	state (\( T=\SI{188}{\K} \), \( P=\SI{9}{\kilo\bar} \)) in the real
	dynamics and its inherent structures.}
	\label{fig:is-vs-rd}
\end{figure}

We evaluated the inherent structures (IS)~\cite{stillinger2015energy} by minimizing the potential energy via the steepest descent algorithm (STEEP) in GROMACS with a force tolerance of \SI{1}{\J\per\mole\per\nano\m} and a maximum step size \SI{5e-4}{\nano\m}. To ensure maximum accuracy, the minimization was carried in double precision.
An example of the effect of the energy minimization on the liquid structure is shown in Fig.~\ref{fig:is-vs-rd}, for \( P=\SI{9}{\kilo\bar} \).
The increase in short-range order causes a significant increase in the height of the first peak, enhancing the separation between the first and second hydration shells. In the intermediate range, only minor differences are observed, related to a slight enhancement of peaks and valleys, while the long-range correlations are practically unaffected.

\begin{figure}
	\centering
	\includegraphics[width=11cm]{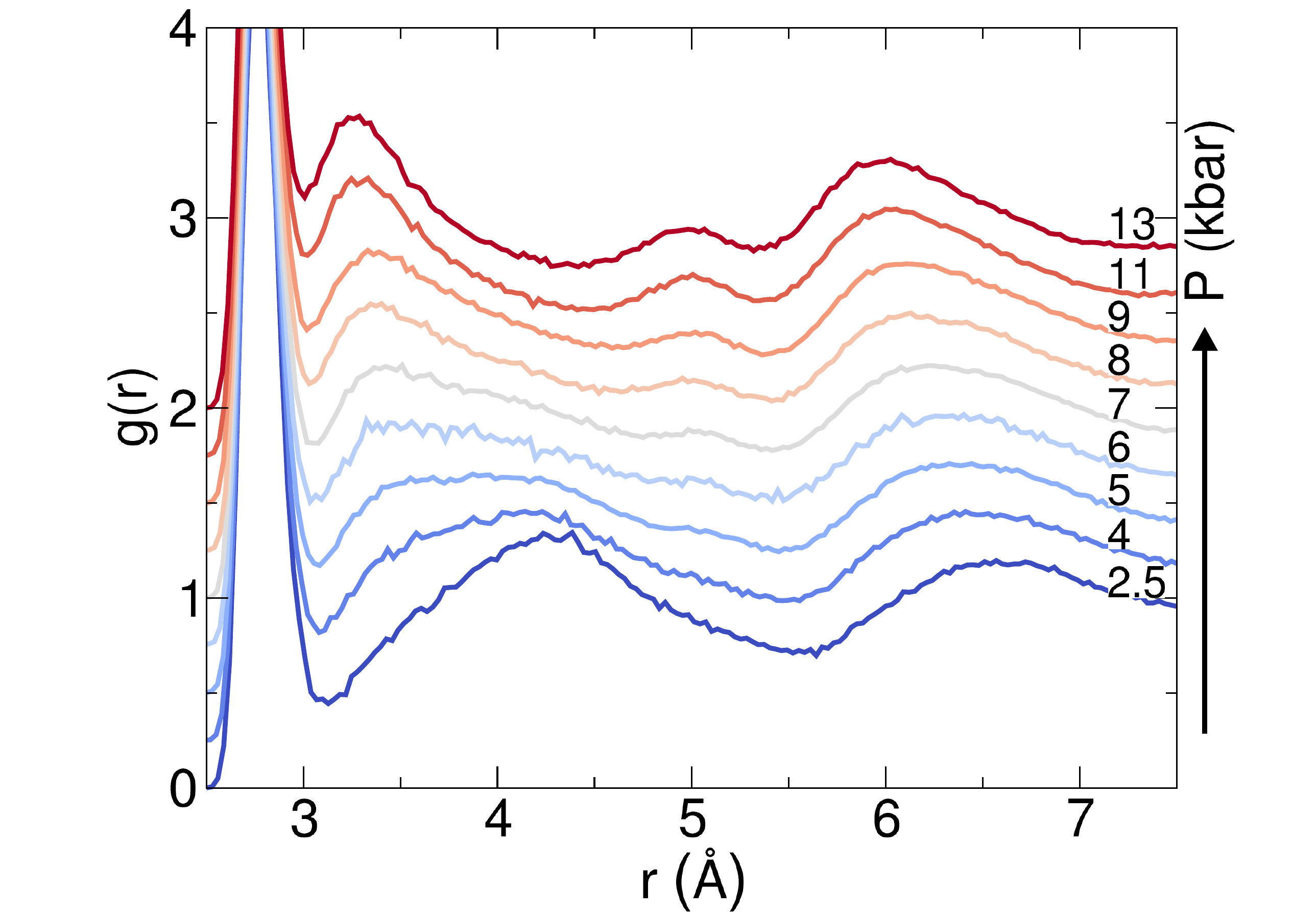}
	\caption{Pressure dependence of the oxygen-oxygen \( g(r) \) evaluated in the real dynamics along the \SI{188}{\K} isotherm. Successive \( g(r) \) curves are shifted by 0.25 along the verical axis.}
	\label{fig:gr_RD}
\end{figure}

In Fig.~\ref{fig:gr_RD} we also show the complete pressure dependence of the \( g(r) \)
in the real dynamics, to be compared with Fig.~2 in main text.
It is evident that the process of energy minimization enchances the structural features of the system without artificially altering them.

\section{Structure factor}
\begin{figure}
	\centering
	\includegraphics[width=11cm]{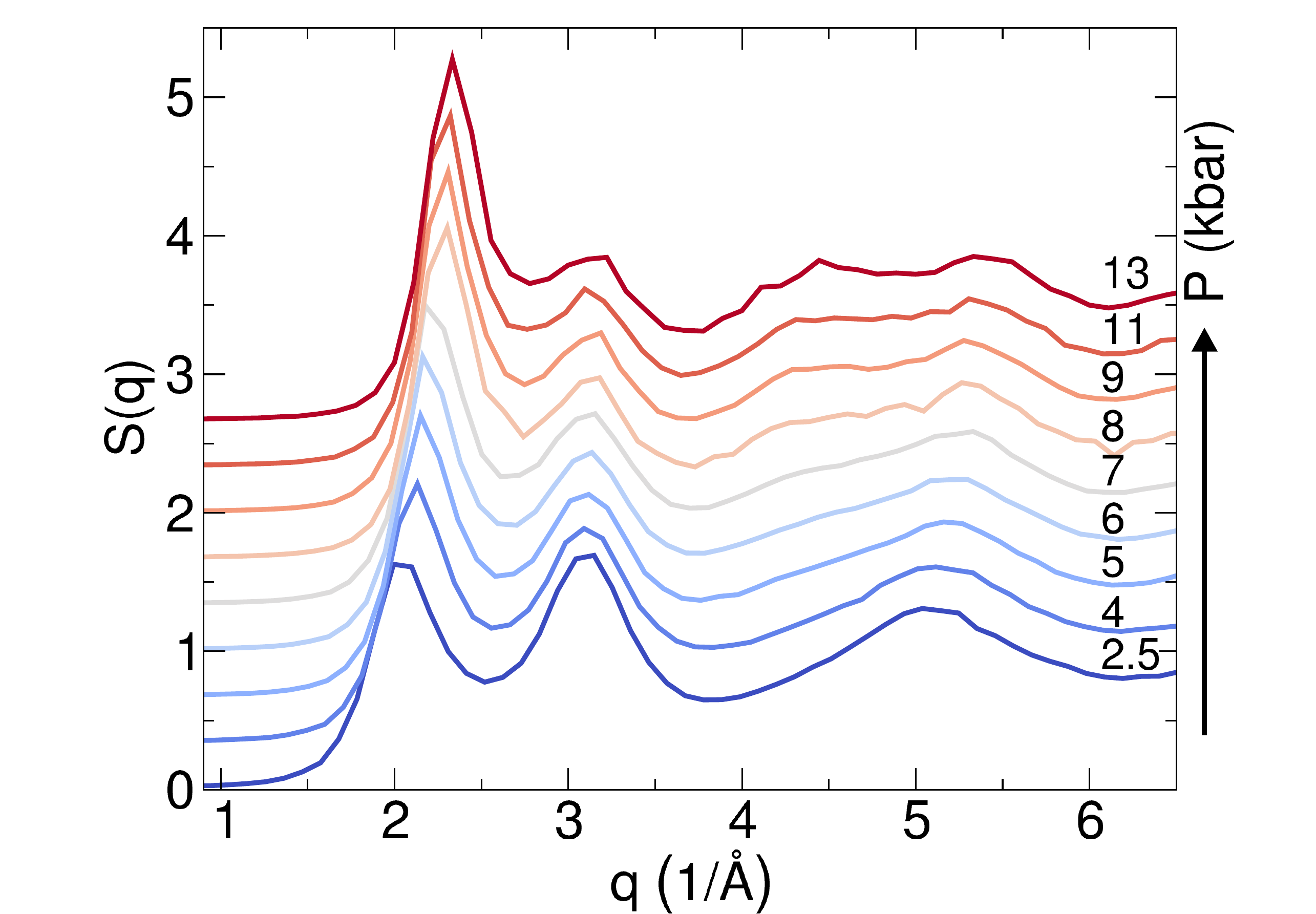}
	\caption{Pressure dependence of the oxygen-oxygen structure factor
	evaluated in the IS along the \SI{188}{\K} isotherm.
	Successive \( S(q) \) curves have been shifted by \( 1/3 \) along the
	vertical axis.}
	\label{fig:sq}
\end{figure}

\begin{figure}
	\centering
	\includegraphics[width=11cm]{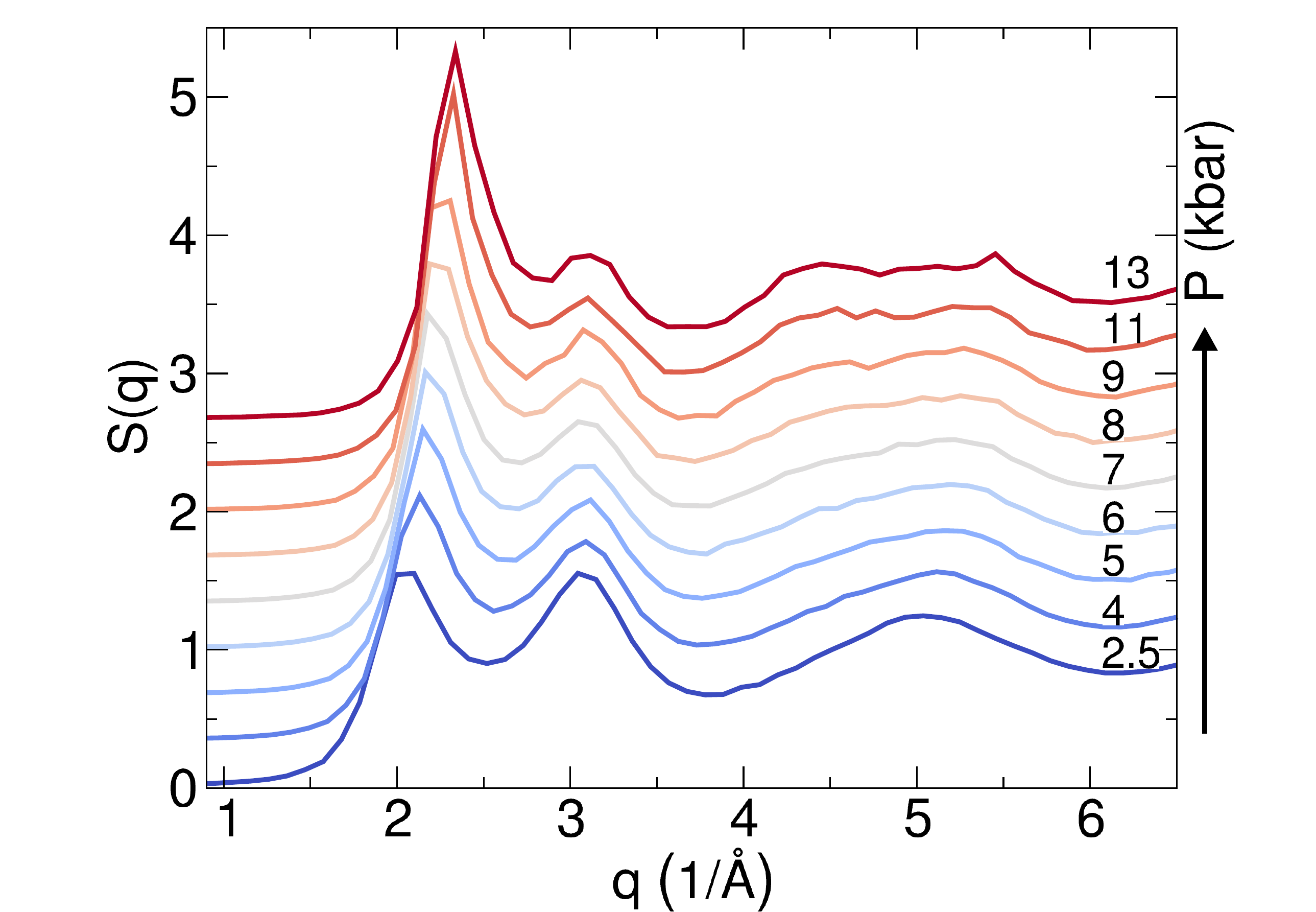}
	\caption{Pressure dependence of the oxygen-oxygen structure factor
	evaluated in the real dynamics along the \SI{188}{\K} isotherm.
	Successive \( S(q) \) curves have been shifted by \( 1/3 \) along the
	vertical axis.}
	\label{fig:sq_RD}
\end{figure}

The structural description provided by the radial distribution function (Fig.~2 in main text) is here complimented by its Fourier transform, the structure factor \( S(q) \)~\cite{hansen2013theory}, reported in Fig.~\ref{fig:sq} (evaluated in the IS).
In Fig.~\ref{fig:sq_RD} \( S(q) \) is evaluated in the real dynamics, and once again we observe that the structural features of the system are not altered by energy minimization.
In Fourier space one observes a clear shift (and enhancement) of the pre-peak from \( \approx 2 \) to \( \approx\SI{2.4}{\AA^{-1}} \) and a decrease in the peak intensity at \( \SI{3.1}{\AA^{-1}} \).
Also, above \SI{7}{\kilo\bar} the peak around \( \SI{5}{\AA^{-1}} \) splits into two broad but distinct peaks, at \( \approx \SI{4.5}{\AA^{-1}} \) and \( \approx \SI{5.5}{\AA^{-1}} \) respectively.
The evolution of these features is consistent with recent experimental observations by \citet{mariedahl2018xray}, increasing our confidence in the ability of TIP4P/Ice to faithfully describe the structural properties of deep supercooled water.

\section{Comparison with experimental data}
To obtain amorphous states that could be compared with the experimental glasses as shown in Fig.~3 of main text, the IS were annealed to \SI{80}{\K} and relaxed to ambient pressure.
The annealing was carried out over an interval of \SI{200}{\pico\s}, and we used the Berendsen algorithm~\cite{berendsen1984molecular} for both temperature and pressure coupling with short relaxation times, respectively 0.8 and \SI{1.8}{\pico\s}. Here we adopted these algorithms to avoid artificial oscillations in $T$ and $P$ which are sometimes observed with other thermostats and barostats after abrupt changes in thermodynamic conditions.
The Berendsen algorithm is known to provide the correct thermodynamics and structural averages, but incorrect fluctuations. 

\begin{figure}
	\centering
	\includegraphics[width=11cm]{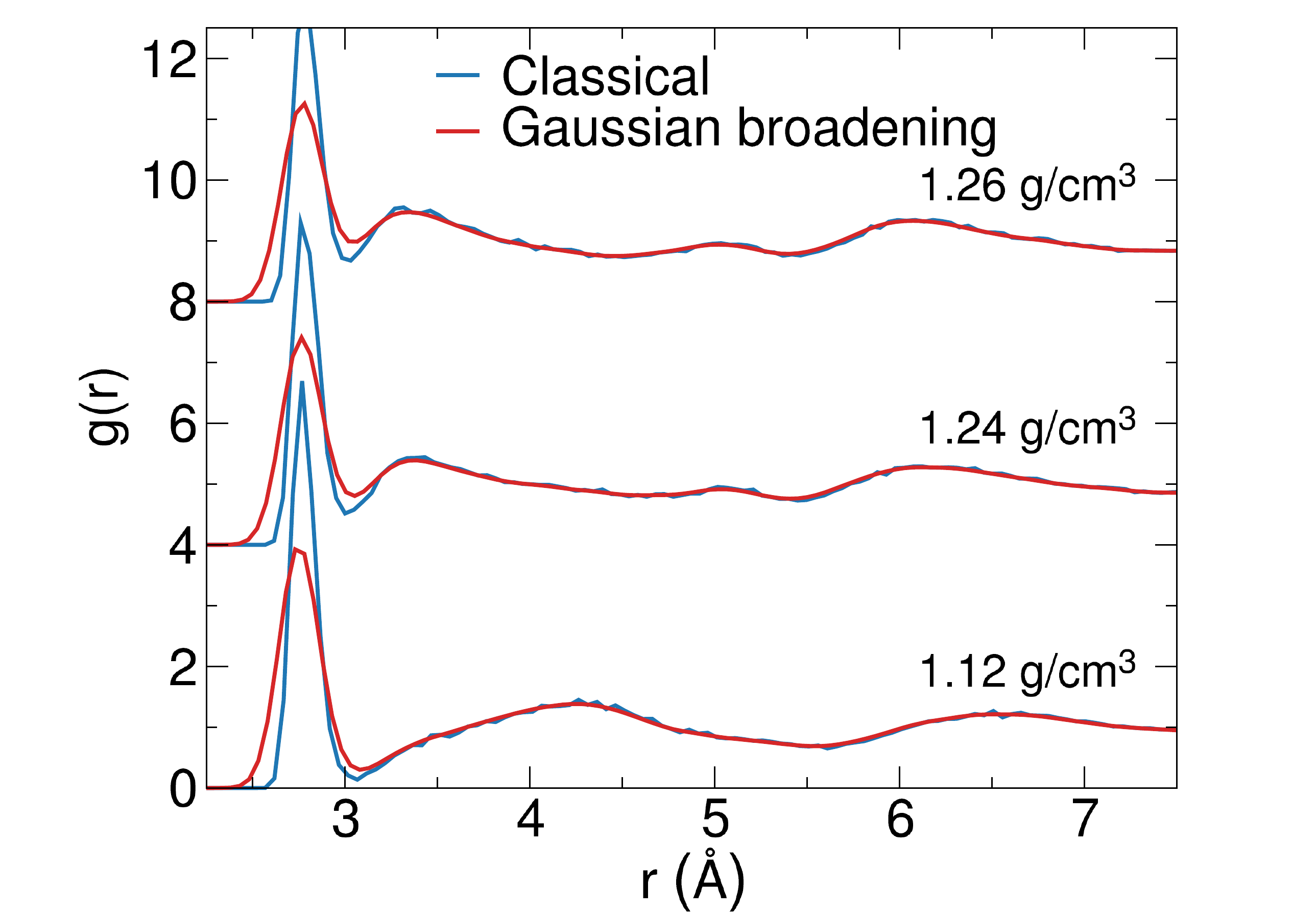}
	\caption{Comparison of numerical \( g(r) \) before and after
	application of quantum broadening corrections, for amorphous
	configurations at three different densities.}
	\label{fig:convolution}
\end{figure}

As explained in main text, the glass structures thus obtained are still not ready to be compared with experimental data; at such low temperatures quantum effects become relevant~\cite{herrero2012highdensity} and a correction, accounting for the delocalization of \ce{O} atoms, has to be applied to the numerical results from the classical TIP4P/Ice potential.
Fig.~3 shows the comparison of numerical and experimental data, after the numerical \( g(r) \) has been  convoluted with a Gaussian of variance \SI{7.1e-3}{\AA\squared} as discussed in the main article; in Fig.~\ref{fig:convolution} we compare the \( g(r) \) before and after such correction is applied, for the glasses recovered at three different values of density.
As expected, the convolution affects only the region of the first peak, while the correlations at larger separations are unaffected (although the curve is obviously smoothened).

\section{Hydrogen bonds}
To perform an accurate analysis based on topological properties 
it is of fundamental importance to correctly identify HBs at all thermodynamic conditions.
Several different definitions of HB have been proposed throughout the years~\cite{kumar2007hydrogen}.

\begin{figure}
	\centering
	\includegraphics[width=16cm]{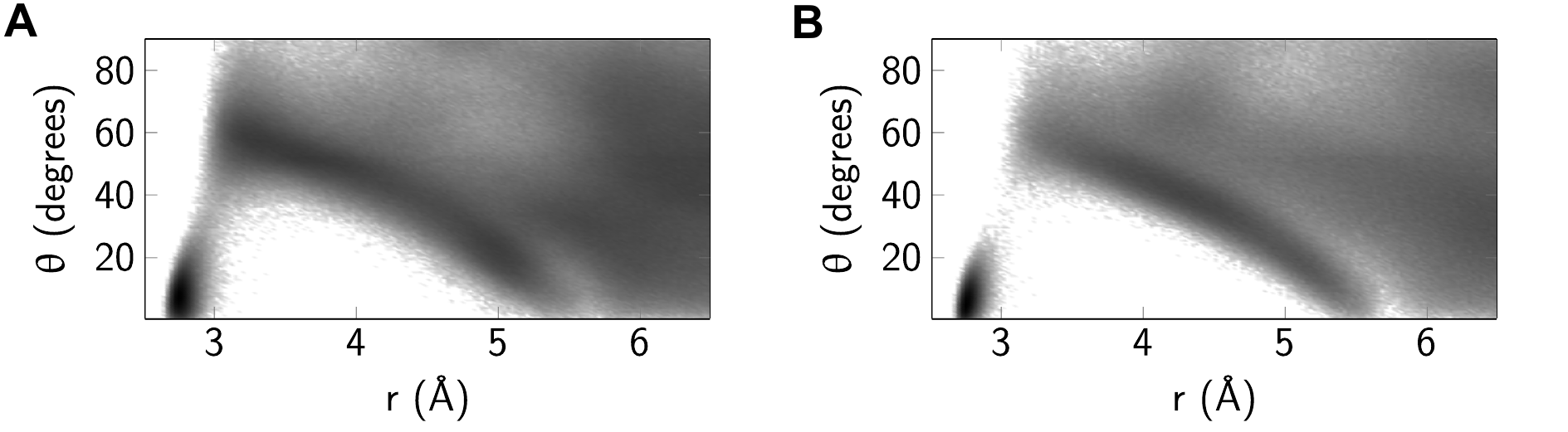}
	\caption{Spatial density function, \( g(r,\theta) \), of the
	\ce{OO} distance \( r \) and the
	\( \ce{H}\hat{\ce{O}}\ce{\bond{...}O} \) angle
	\( \theta \) at (a) \SI{13}{\kilo\bar} and (b) \SI{2.5}{\kilo\bar}.
	The value of \( g(r,\theta) \) is represented by 
	intensity in logarithmic scale, increasing from white to black.}
	\label{fig:bondmaps}
\end{figure}

The definition of \citet{luzar1996hydrogenbond} (LC)
(\( r_\text{OO} < \SI{3.5}{\angstrom} \) and \( \theta < \SI{30}{\degree} \) where \( \theta \) is the  (smallest) \( \ce{H}\hat{\ce{O}}\ce{\bond{...}O} \) angle) when applied to the IS allows us to identify HBs with high accuracy, despite the increasing distortion in the local environments that is observed upon isothermal compression. Figure~\ref{fig:bondmaps} depicts the \( g(r,\theta) \) distribution of the system at 2.5 and \SI{13}{\kilo\bar}.
With increasing pressure, the identification of HBs provided by the LC definition becomes less accurate since the distribution around the main HB peak (\( r\approx\SI{2.8}{\angstrom} \), \( \theta\approx\SI{5}{\degree} \)) becomes broader. It should be noticed, however, that point density in Fig.~\ref{fig:bondmaps} is shown in log scale, so that low-probability regions in \( r-\theta \) space are accentuated with respect to high-probability ones.

\begin{figure}
	\centering
	\includegraphics[width=11cm]{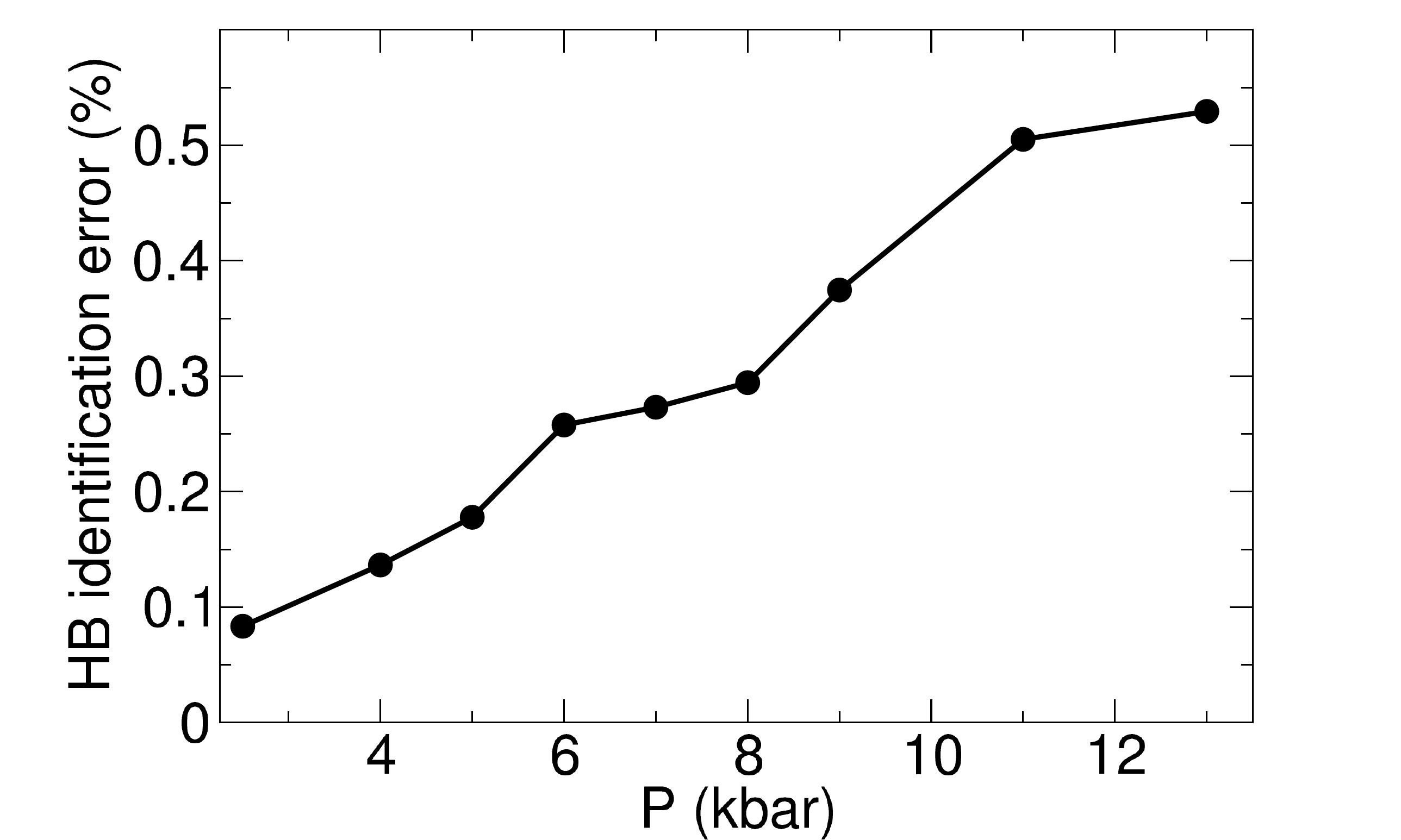}
	\caption{Average identification error of HBs as a function of pressure
	with the Luzar-Chandler definition.}
	\label{fig:hberror}
\end{figure}

The average identification error committed by adopting the LC definition can be estimated as follows.
We can consider ``proper'' HBs those that satisfy the LC conditions, and ``uncertain'' HBs those with a bond length \( r < \SI{3.5}{\angstrom} \) but with an angle within 20\% above the LC angular threshold:
\( \SI{30}{\degree} < \theta < \SI{36}{\degree} \).
It appears likely that at least a fraction of these  uncertain bonds should be considered as proper HBs.
The error can then be defined as the ratio between the number of uncertain and proper HBs.
The results are reported in Fig.~\ref{fig:hberror}, showing that the average error is below 1\% even at the highest pressure.
The LC definition of HB is therefore reasonably valid over the whole isotherm.

\begin{figure}
	\centering
	\includegraphics[width=11cm]{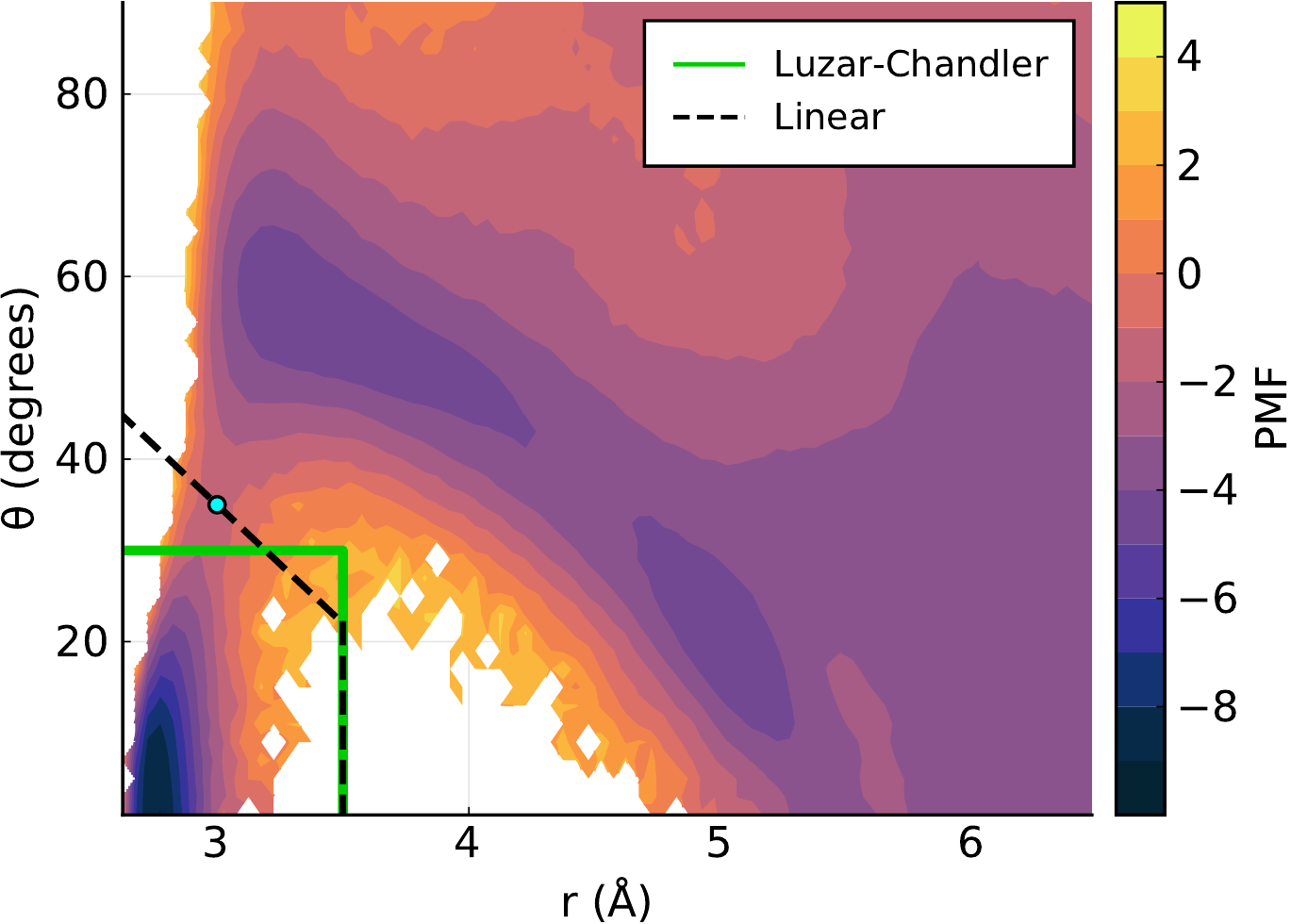}
	\caption{Comparison, at \( P=\SI{13}{\kilo\bar} \), of the
	Luzar-Chandler definition of HB and the new
	definition obtained by evaluating the linear path
	(\( \theta=\theta_0+\alpha r \)) of maximum PMF through the saddle
	point (cyan circle, found approximately at
	\( r\approx\SI{3}{\angstrom} \) and
	\( \theta\approx\SI{35}{\degree} \)).}
	\label{fig:newhb}
\end{figure}

To investigate an alternative definition,  instead of using the LC angular threshold \( \theta^*=\SI{30}{\degree} \),
we  choose a \( r \)-dependent angular threshold \( \theta^*=\theta^*(r) \)
with the aim of minimizing the error intrinsic in the definition.  
The simplest approach uses a linear dependence, \( \theta^*(r) = \theta_0 + \alpha r \).
We thus identify a potential of mean force $PMF$ (per unit thermal energy)
\begin{equation}
	\text{PMF} = -\ln g(r,\theta)
\end{equation}
and obtain the values of \( \theta_0 \) and \( \alpha \) by identifying the linear path of maximum PMF passing through the saddle point at \( r\approx\SI{3}{\angstrom} \) and \( \theta\approx\SI{35}{\degree} \).
This path identifies the optimal boundary between H-bonded and non-H-bonded pairs. By optimising over the data at \( P=\SI{13}{\kilo\bar} \), we find:
\begin{equation}
	\theta_0 \approx \SI{113}{\degree}, \qquad
	\alpha \approx \SI{-26}{\degree\per\angstrom}.
\end{equation}

With this method, two molecules are H-bonded if their \ce{O-O} distance is \( r<\SI{3.5}{\angstrom} \) and their \( \ce{H}\hat{\ce{O}}\ce{\bond{...}O} \) angle satisfies
\( \theta < \SI{113}{\degree} - \SI{26}{\degree\per\angstrom}r \). See Fig.~\ref{fig:newhb}. At \( P=\SI{13}{\kilo\bar} \), where the difference between the two definitions is largest, the new definition counts an average of \( \approx 2002 \) H-bonded pairs out of 1000 molecules, whereas the LC criterion identifies \( \approx 1995 \) pairs; that is, an average difference of 7 HBs per configuration.
As the pressure is decreased, the difference between the two definitions  decreases.
Given the minute difference, we conclude that even in this high-\( P \) regime, the LC definition can be adopted without the risk of significantly compromising the topological description of the system. Hydrogen bonds in the following (and in the main text)  are thus always identified according to the LC definition in the IS.

\section{Network defects}
In the explored thermodynamic range, we observe 
molecules  with coordination numbers 3 or 5 (network defects); 
Following the nomenclature introduced by~\citet{saito2018crucial}, 
5-coordinated defects donate 2 HBs and accept 3 (\ce{H^2O^3}), while 3-coordinated defects can either donate 2 HBs and accept 1 (\ce{H^2O^1}) or donate 1 and accept 2 (\ce{H^1O^2}).
In a previous work~\cite{foffi2021structural} we have shown that network defects appear to have a role in the structural transformations that characterize the liquid-liquid phase transition of (TIP4P/Ice) water;
along the \SI{188}{\K} isotherm, from ambient pressure up to \SI{2.5}{\kilo\bar}, \ce{H^2O^1} and \ce{H^2O^3} were found to be almost-perfectly paired, thus conserving the total number of HBs and imposing only a minor energetic cost on the system; the presence of \ce{H^1O^2}, a geometry that sacrifices a HB, was negligible.

\begin{figure}
	\centering
	\includegraphics[width=11cm]{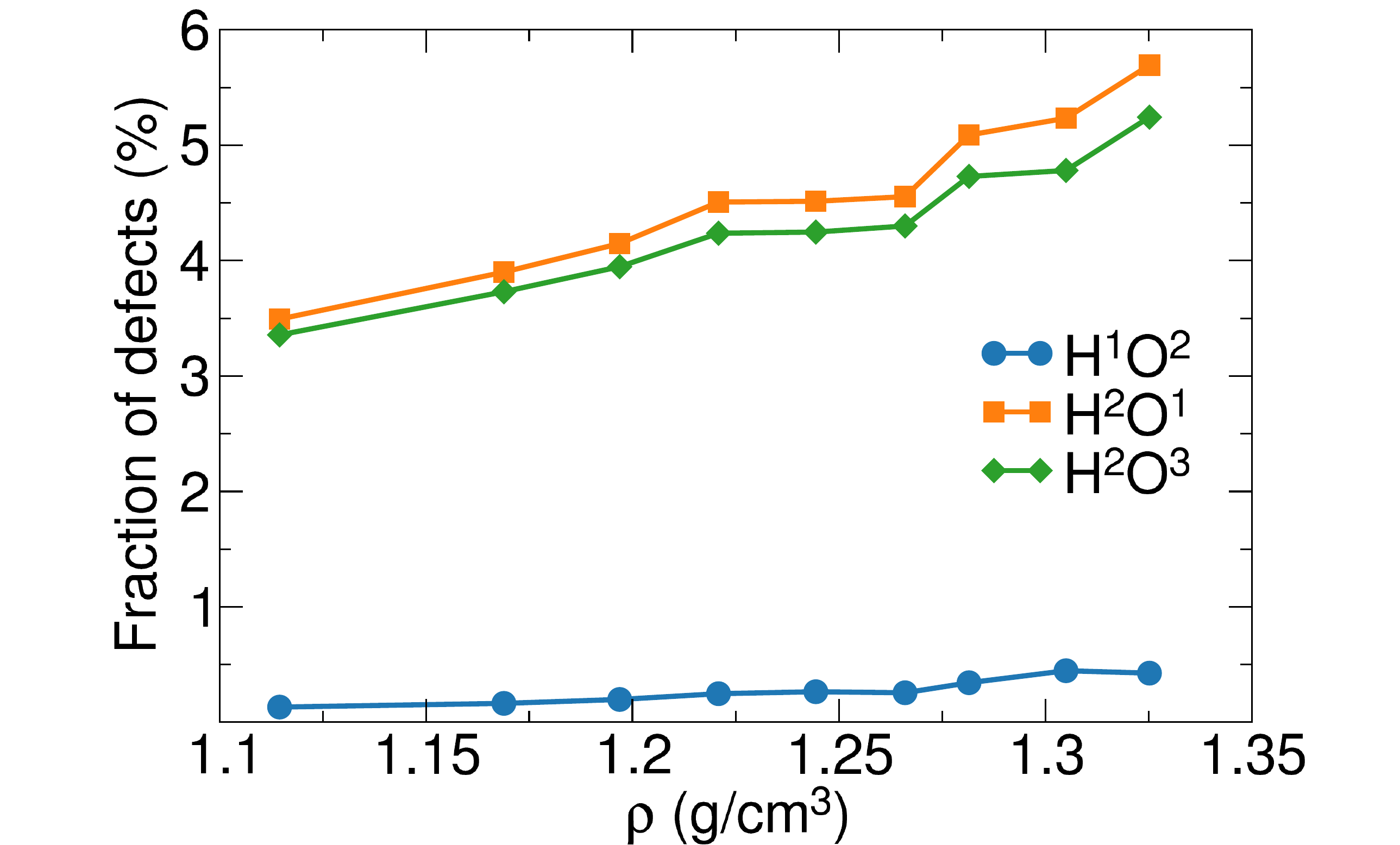}
	\caption{Population of network defects as a function of density.}
	\label{fig:defects-vs-rho}
\end{figure}

The fraction of defects, distinguished by species, that we observed along 
the \SI{188}{\K} isotherm from 2.5 to \SI{13}{\kilo\bar} is shown in 
Fig.~\ref{fig:defects-vs-rho}.
At all pressures, more than 99.5\% of the H atoms are involved in HBs, 
indicating that, despite the high pressure and intense structural 
distortion, practically all molecules donate two hydrogens for bonding.
While most molecules still participate in four HBs, there is a  
fraction of network defects, accepting either one or three bonds, whose 
concentration increases (roughly) linearly with density.
The strong correlation between the number of five- and three-coordinated 
molecules and the density might suggest that the primary effect of 
compression is to convert a pair of tetrahedrally-coordinated molecules 
into a \ce{H^2O^1 - H^2O^3} defect pair. 
Consistent with the weak dependence of $E$ on $\rho$ (see Fig.~1 of the main text),
such a transformation does not have a significant energetic cost, 
preserving the total number of HBs while improving the packing.

\begin{figure}
	\centering
	\includegraphics[width=11cm]{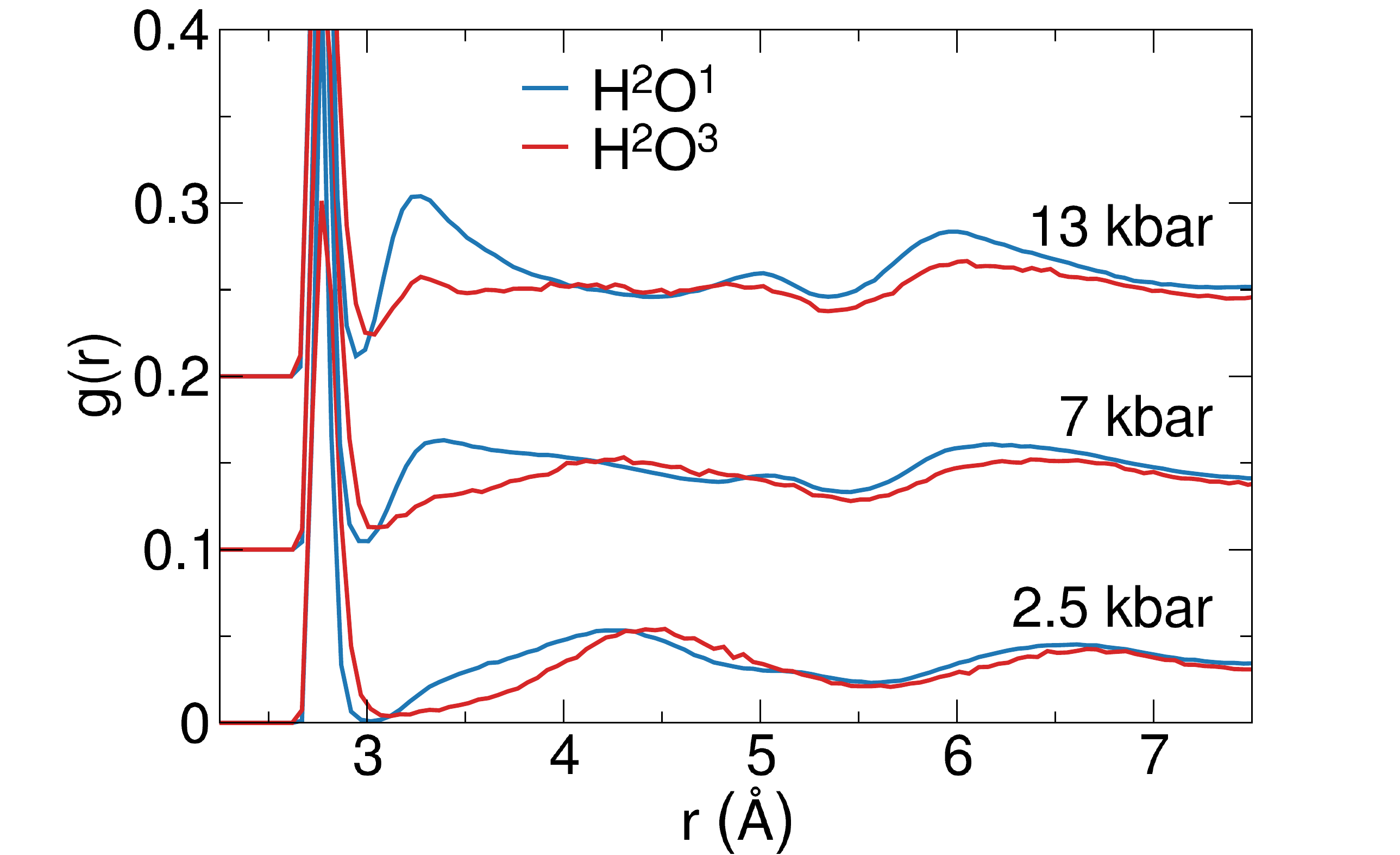}
	\caption{Contribution of network defects to \( g(r) \) evaluated in
	the IS at three different pressures along the \SI{188}{\K} isotherm.
	Curves were shifted by 0.1 along the vertical axis;
	the \( r\to\infty \) limit indicates the relative concentration of the
	associated defect species.}
	\label{fig:rdfdefects}
\end{figure}

The structural properties of network defects in this thermodynamic regime are also consistent with the observations  made in Ref.~\cite{foffi2021structural}.
Figure~\ref{fig:rdfdefects} shows the contribution to the \( g(r) \) from network defects at three different pressures; the two curves represent the radial structure around \ce{H^2O^1} and \ce{H^2O^3} molecules, respectively. It is therefore clear that 3-coordinated molecules participate actively in the interpenetration phenomenon, while 5-coordinated ones display
an increasingly featureless distribution as \( P \) increases.

\section{Topology and structure}

\begin{figure}
	\centering
	\includegraphics[width=11cm]{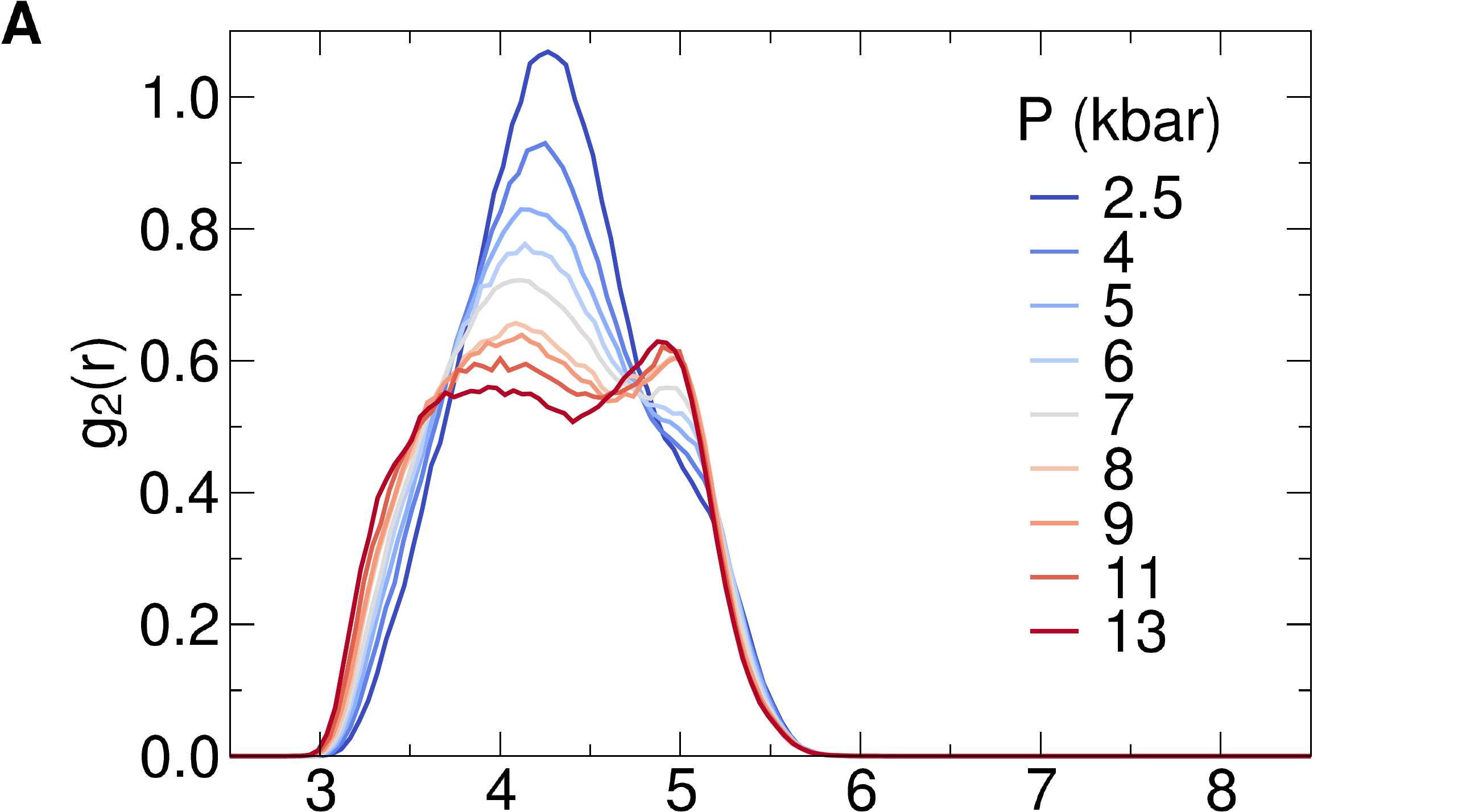}
	\includegraphics[width=11cm]{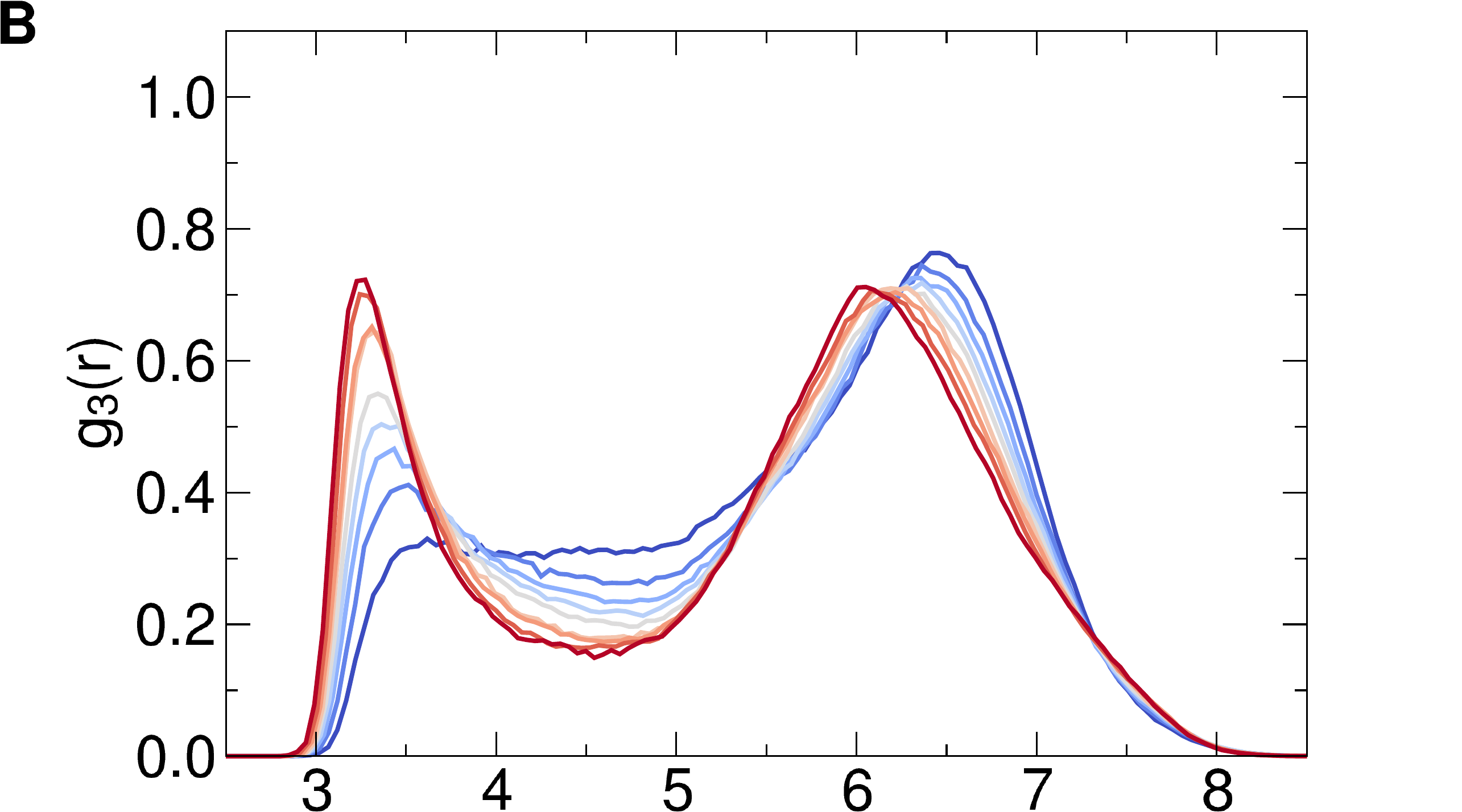}
	\includegraphics[width=11cm]{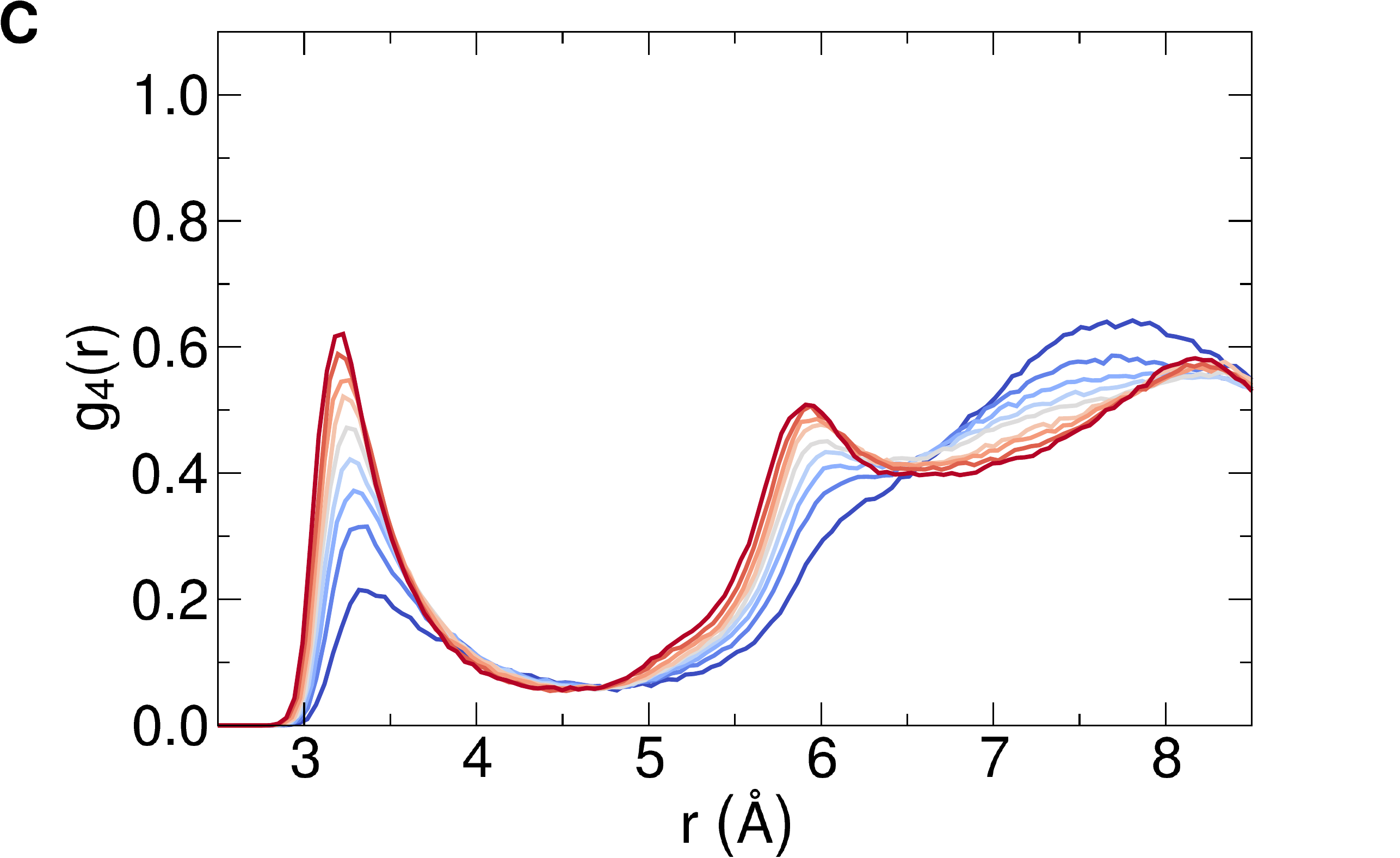}
	\caption{Contributions to the \( g(r) \) along the \SI{188}{\K}
	isotherm separated by coordination shells:
	(a) \( D=2 \), (b) \( D=3 \), (c) \( D=4 \).
	Pressure increases from blue to red.}
	\label{fig:grchemdist}
\end{figure}

The core results about the structure of the system are obtained by classifying the pairs of molecules in the system according to their chemical distance and ring length, as outlined in the main text.
The definition of ring that we adopted here is the same that we proposed in Ref.~\cite{foffi2021structural}. For a given pair of molecules \( (i,j) \) at chemical distance \( D \), the ring length \( L \) associated to this pair is the length (measured in steps along the HB network) of the closed path obtained by joining the two shortest non-intersecting HB-paths between \( i \) and \( j \). By definition, \( L\geq 2D \).

With this approach, we first separate all pairs of molecules in the system depending on their chemical distances \( D \), so as to evaluate the contribution to the radial distribution function \( g(r) \) arising  from  pairs at chemical distance \( D \), such that
\begin{equation*}
	g(r) = \sum_D g_D(r).
\end{equation*}
The results of this decomposition at pressures 2.5 and \SI{13}{\kilo\bar} are shown in Fig.~4 of the main text; here, in Fig.~\ref{fig:grchemdist} we report the complete pressure dependence of the \( g_D \) contributions (for \( D = 2,3,4 \)) along the \SI{188}{\K} isotherm.

As a second step, each pair of molecules at chemical distance \( D \) is further characterized in terms of its ring length \( L \); the total structure of the system is now decomposed in such a way that
\begin{equation*}
	g(r) = \sum_D \sum_L g_D^L(r).
\end{equation*}

\begin{figure}[htbp]
	\centering
	\includegraphics[width=11cm]{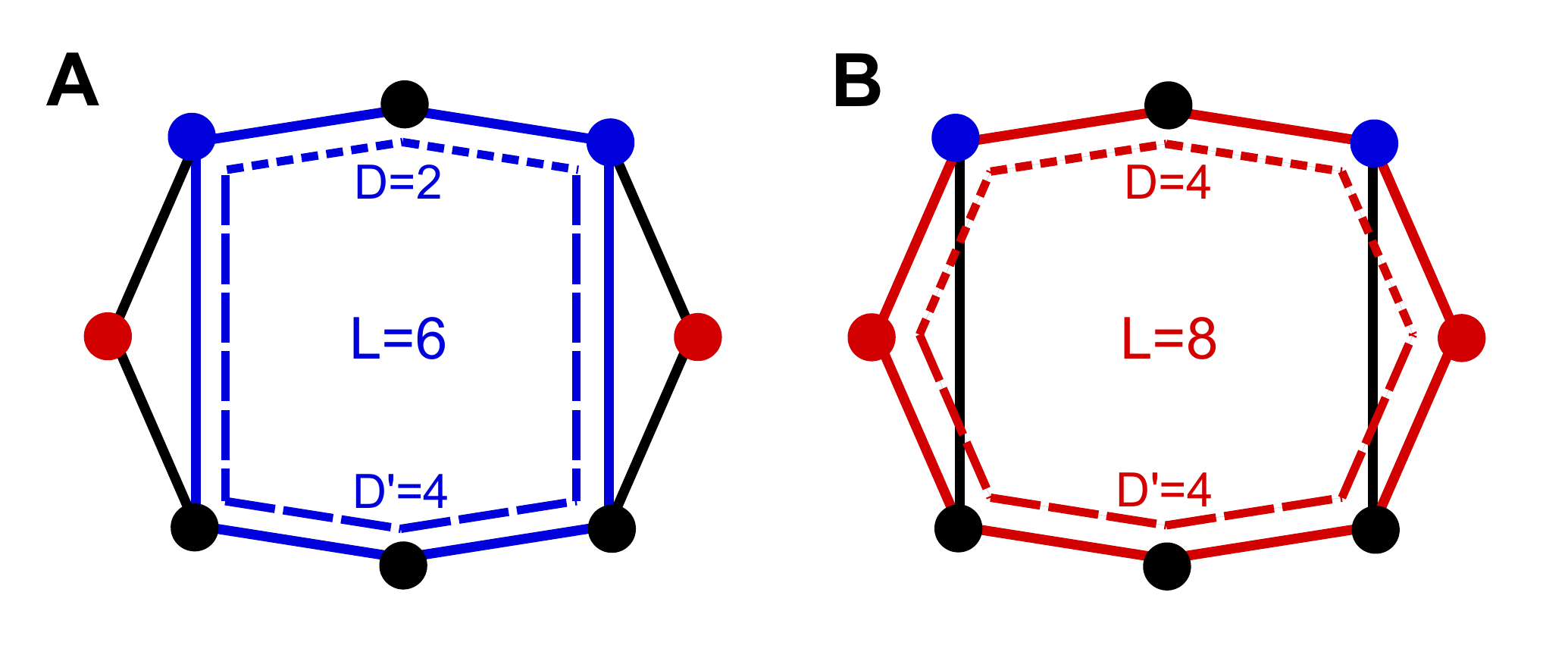}
	\caption{Sample schematics of a HB network showing how rings and ring lengths are identified. (a) The blue molecules are at chemical distance \( D=2 \) (short-dashed path), and the second-shortest path (long-dashed) has length \( D'=4 \); by joining these two paths we obtain a ring of length \( L=6 \). (b) The two shortest paths between the pair of red molecules are both of length 4, producing a ring of length \( L=8 \).}
	\label{fig:howtorings}
\end{figure}

Fig.~\ref{fig:howtorings} shows a sample HB network, where two different pairs of molecules are highlighted, in blue and red respectively.
The ring length associated to the blue pair (a) is \( L=6 \), and that associated to the read pair (b) is \( L=8 \); the two pairs are characterized by different ring lengths, despite the fact that the blue molecules are included in the ring that connects the red pair.

\begin{figure}
	\includegraphics[width=8cm]{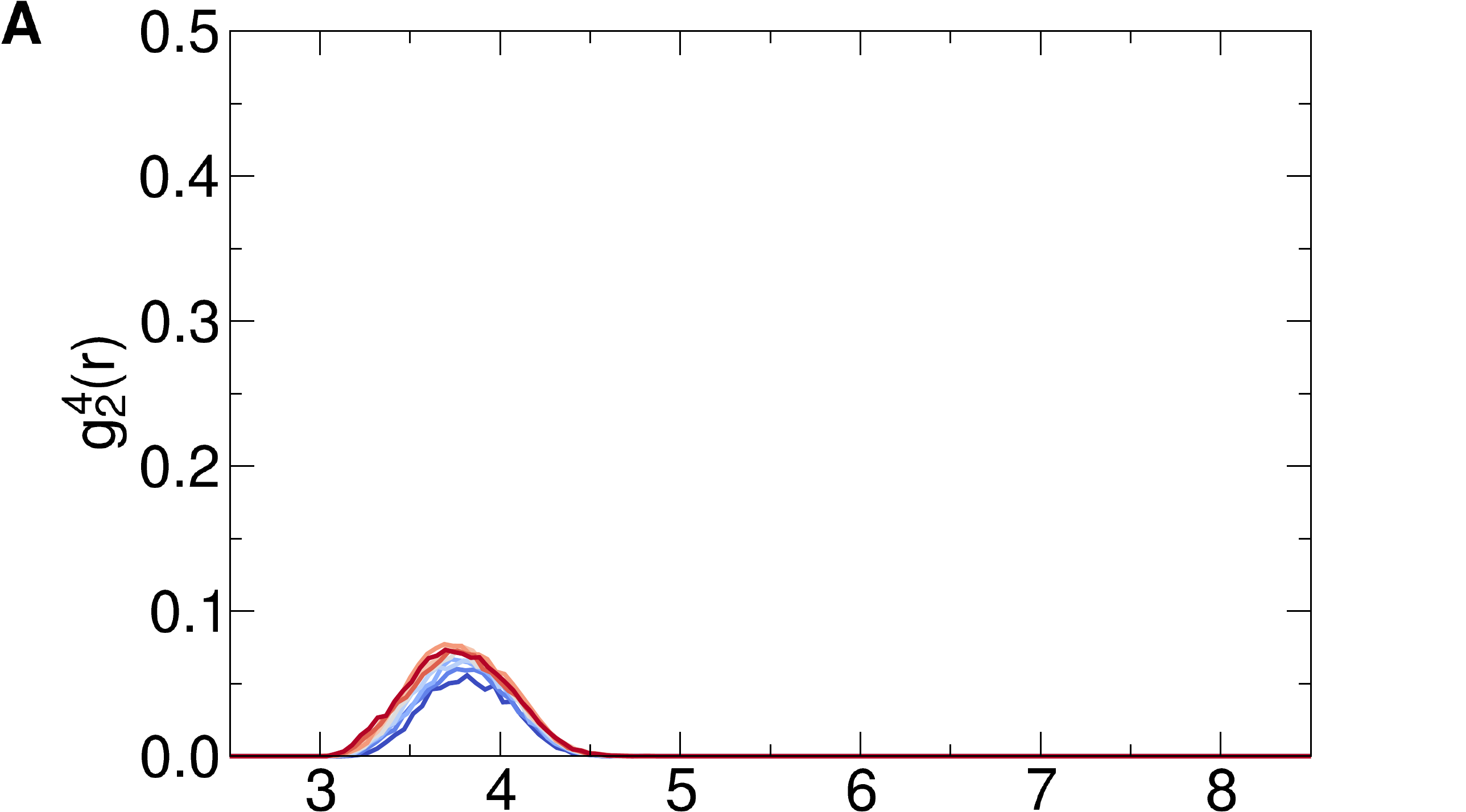}
	\includegraphics[width=8cm]{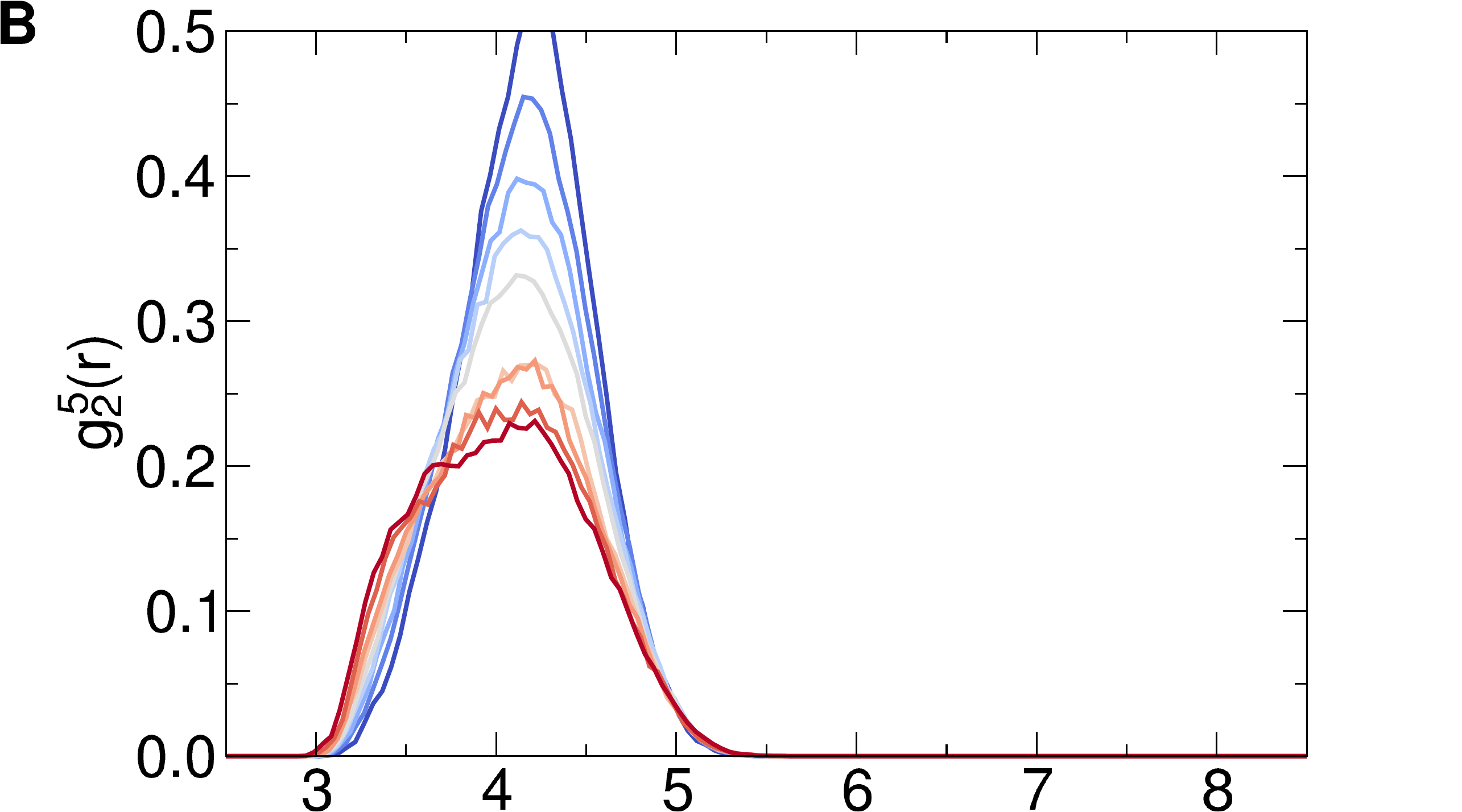}
	\includegraphics[width=8cm]{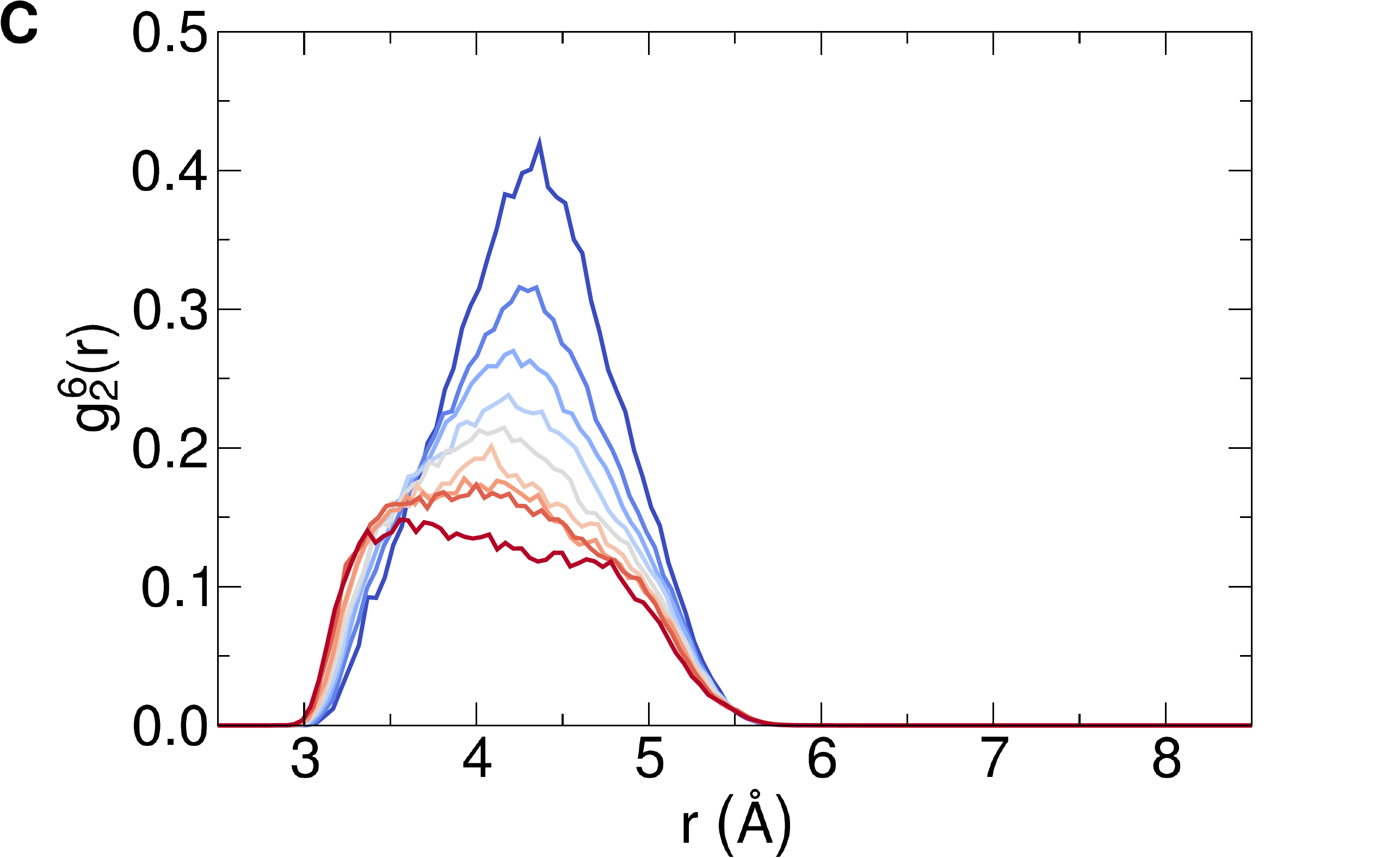}
	\includegraphics[width=8cm]{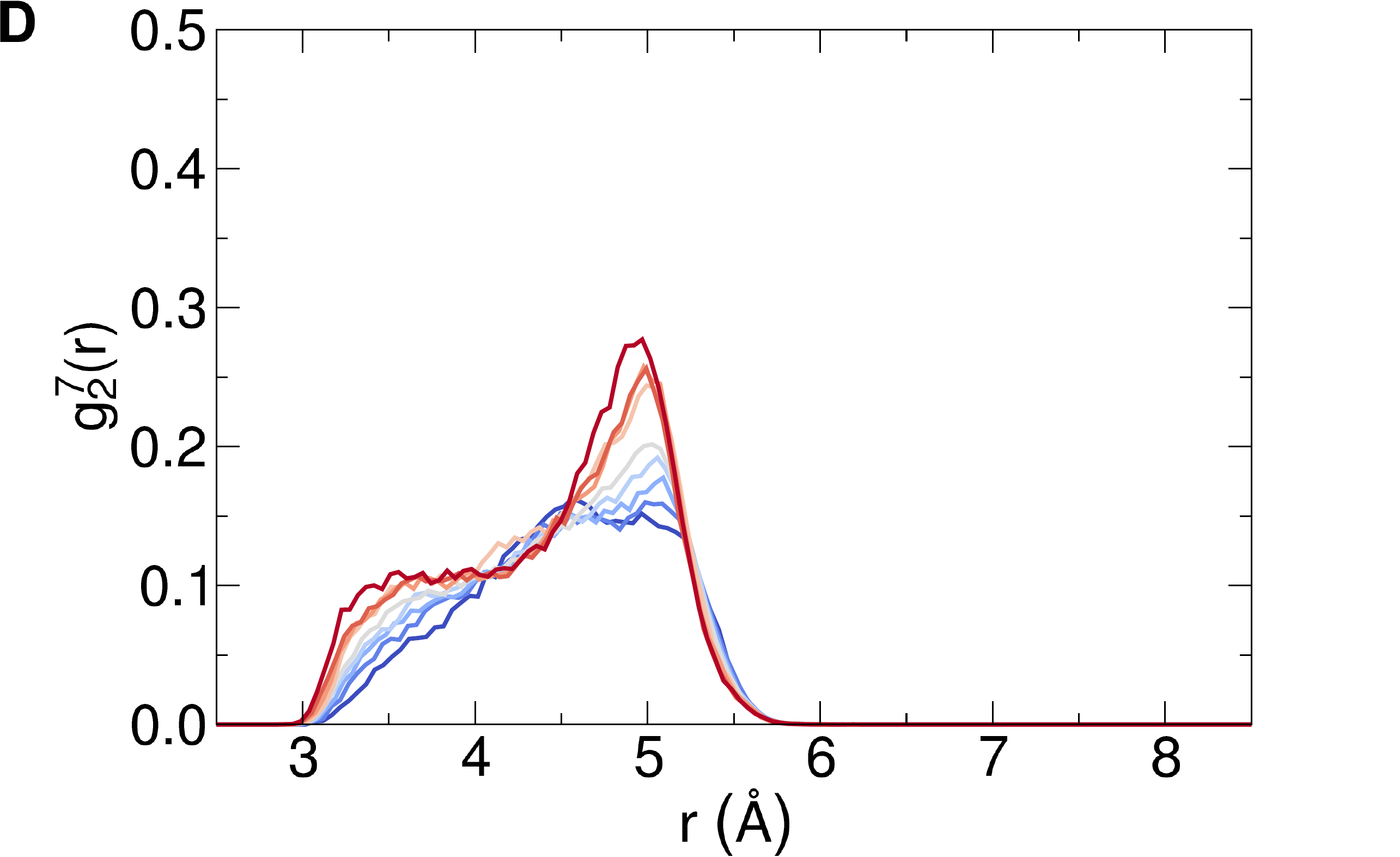}
	\centering
	\vskip 6pt
	\includegraphics[width=8cm]{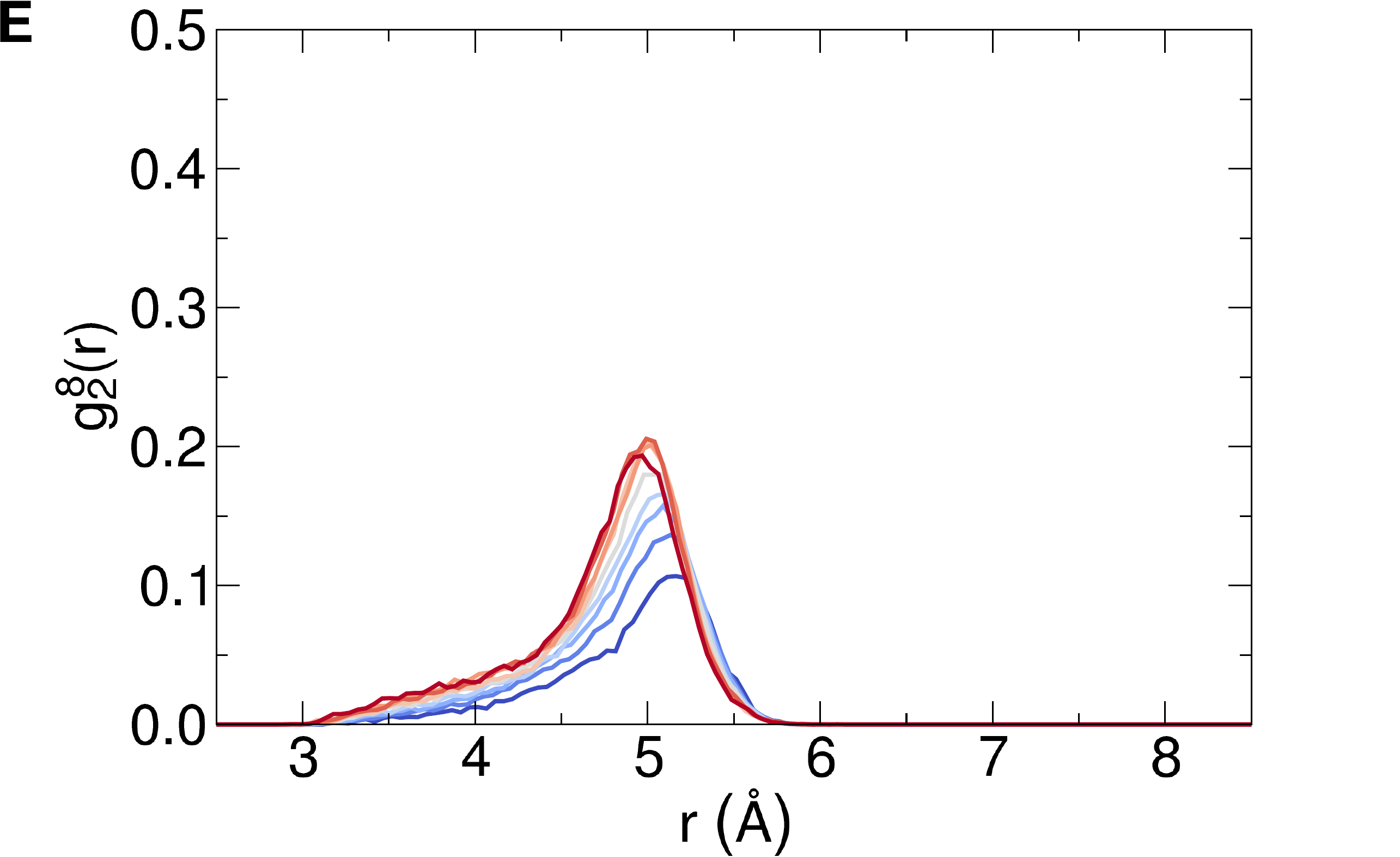}
	\caption{Contributions to the \( g(r) \) arising from pairs of
	molecules at chemical distance \( D=2 \) in rings of length
	\( L \) ranging from (a) \( L=4 \) to (e) \( L=8 \).
	Pressure increases from blue to red.}
	\label{fig:grringsD2}
\end{figure}

\begin{figure}
	\includegraphics[width=8cm]{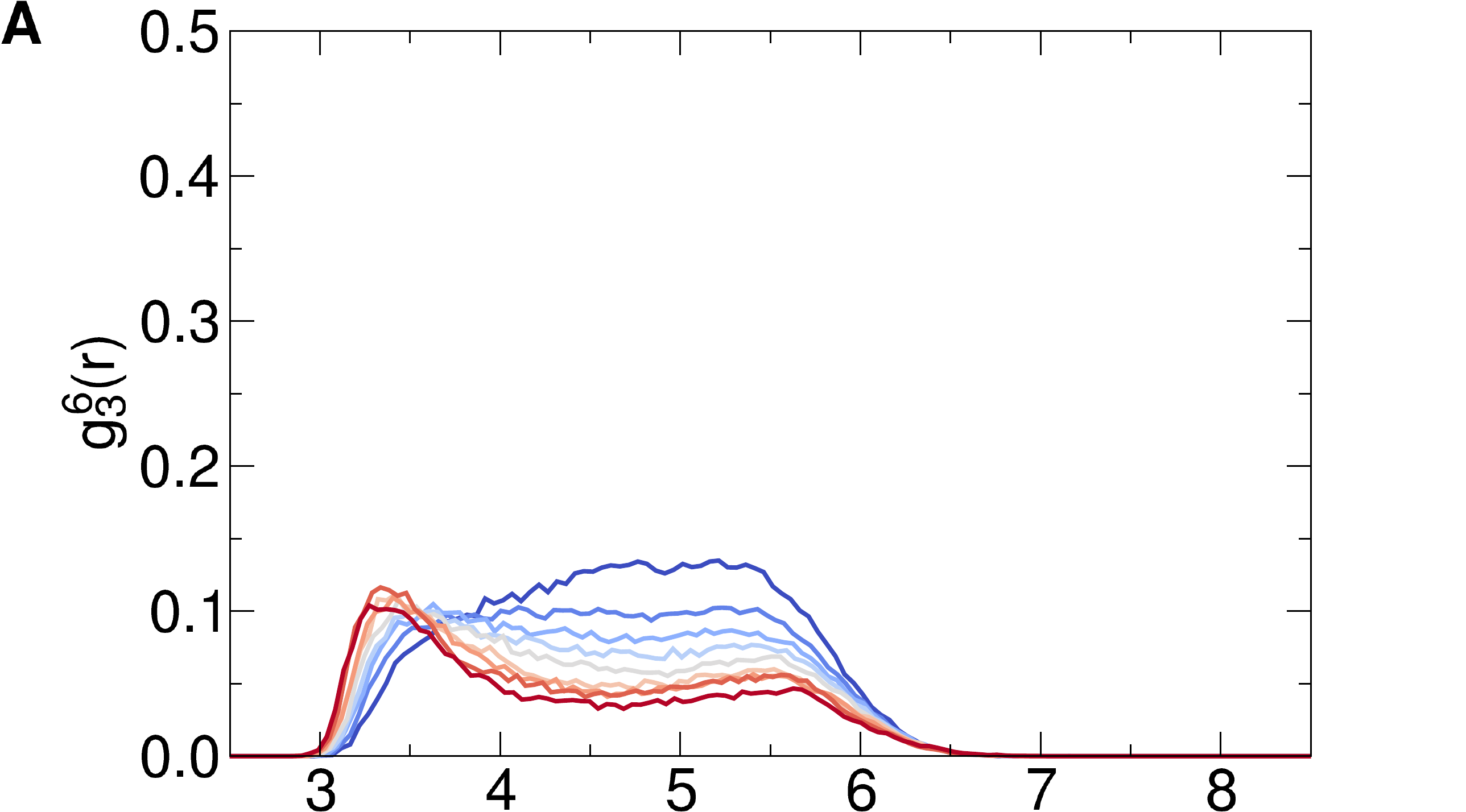}
	\includegraphics[width=8cm]{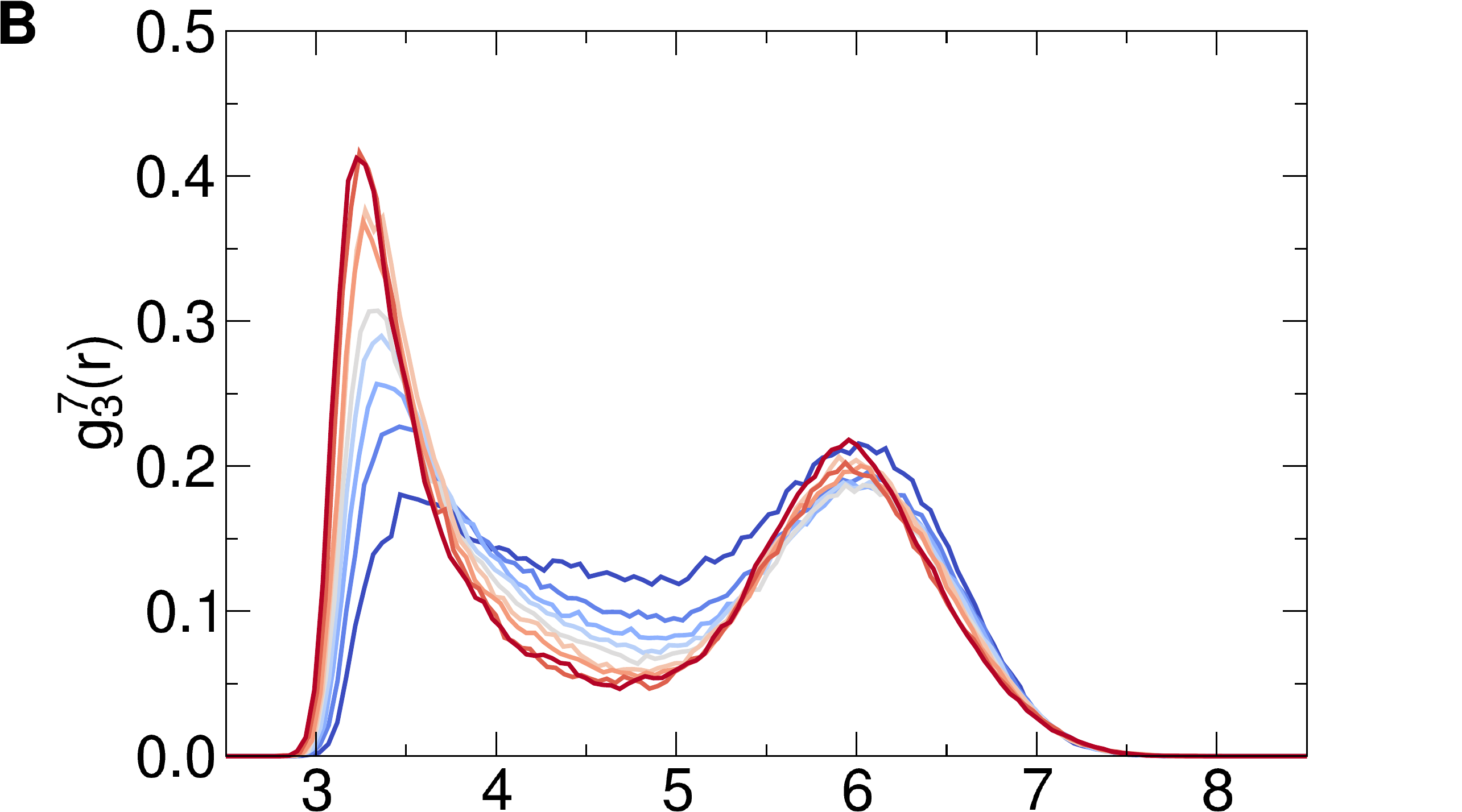}
	\includegraphics[width=8cm]{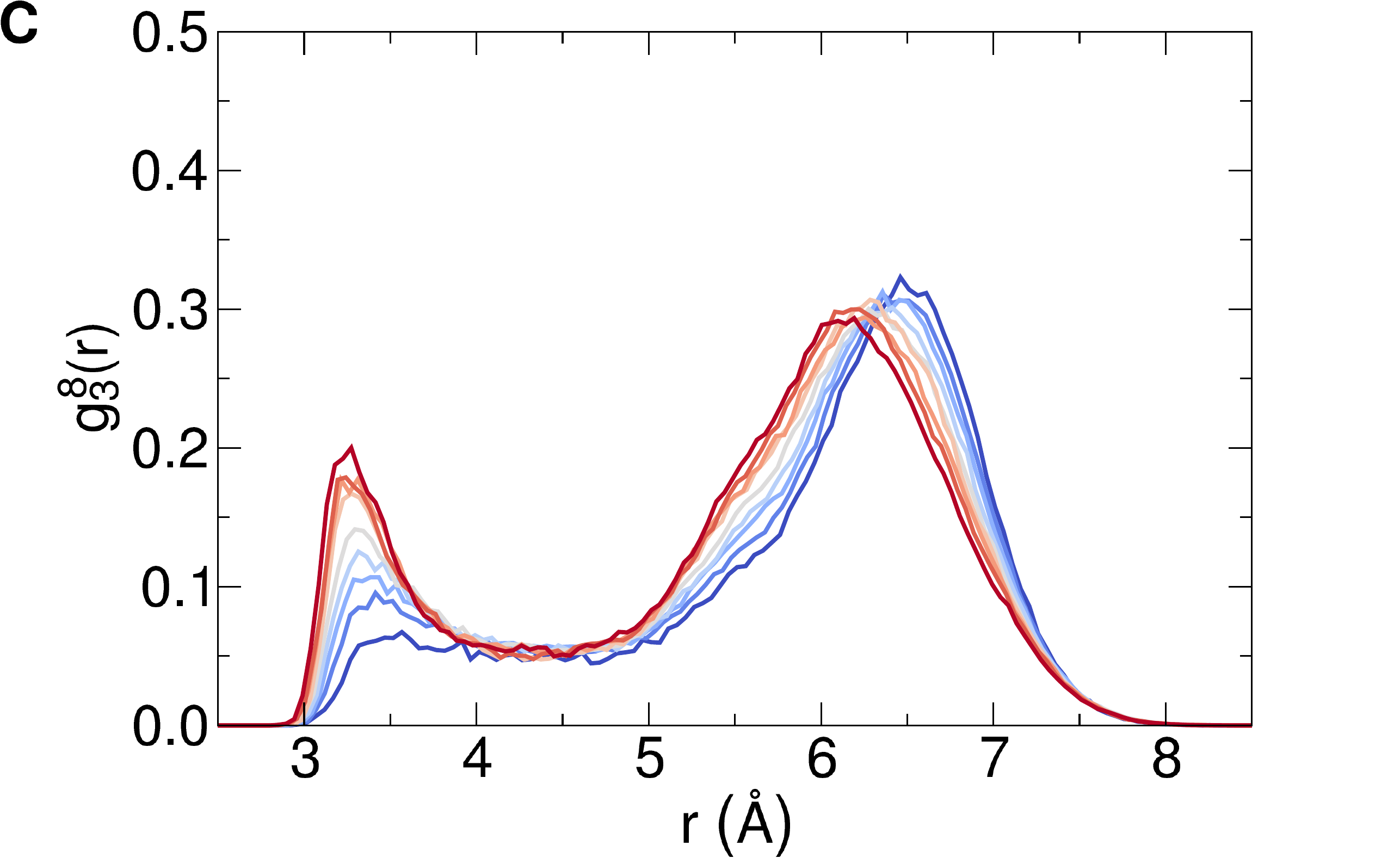}
	\includegraphics[width=8cm]{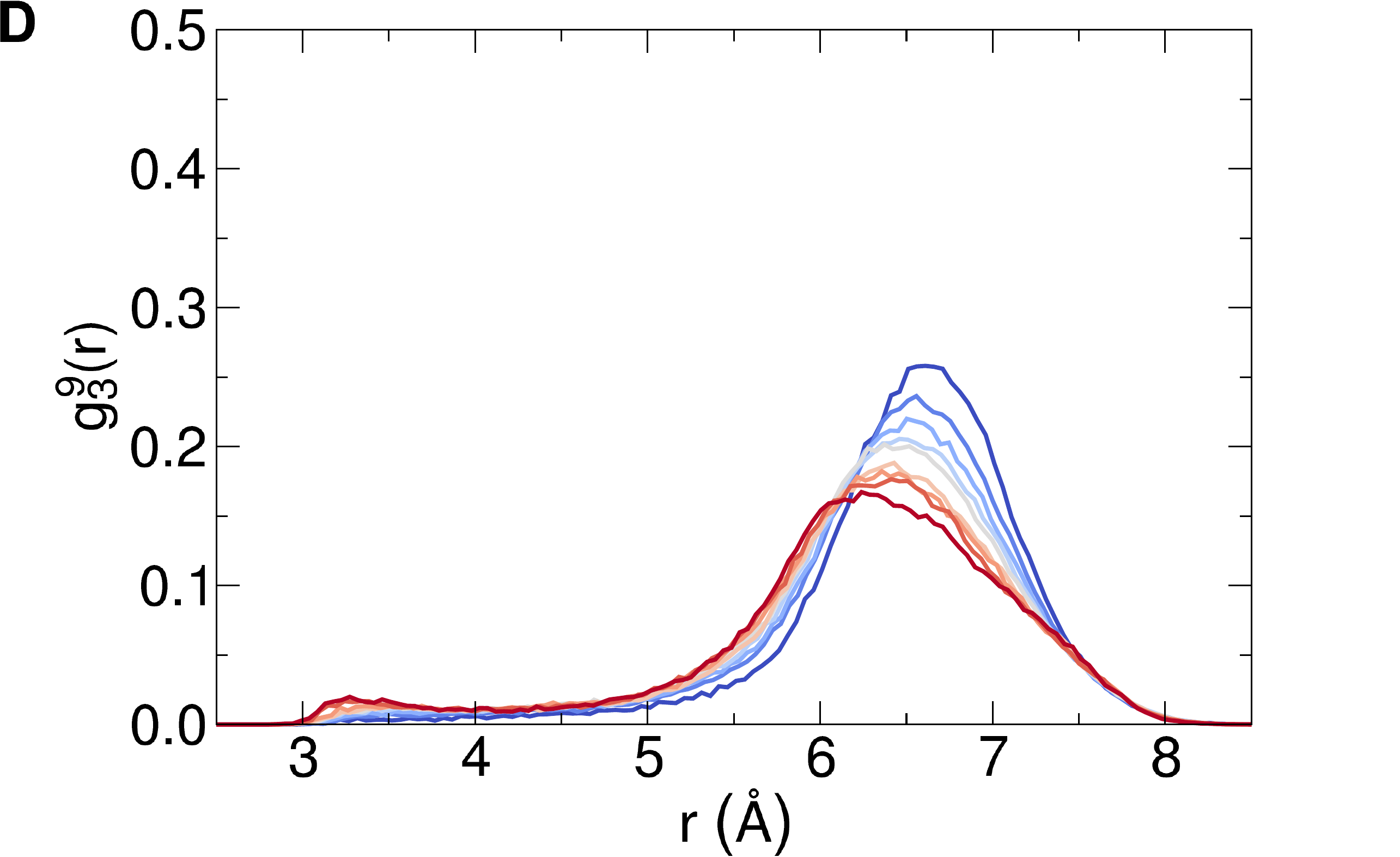}
	\caption{Contributions to the \( g(r) \) arising from pairs of
	molecules at chemical distance \( D=3 \) in rings of length
	\( L \) ranging from (a) \( L=6 \) to (d) \( L=9 \).
	Pressure increases from blue to red.}
	\label{fig:grringsD3}
\end{figure}

\begin{figure}
	\includegraphics[width=8cm]{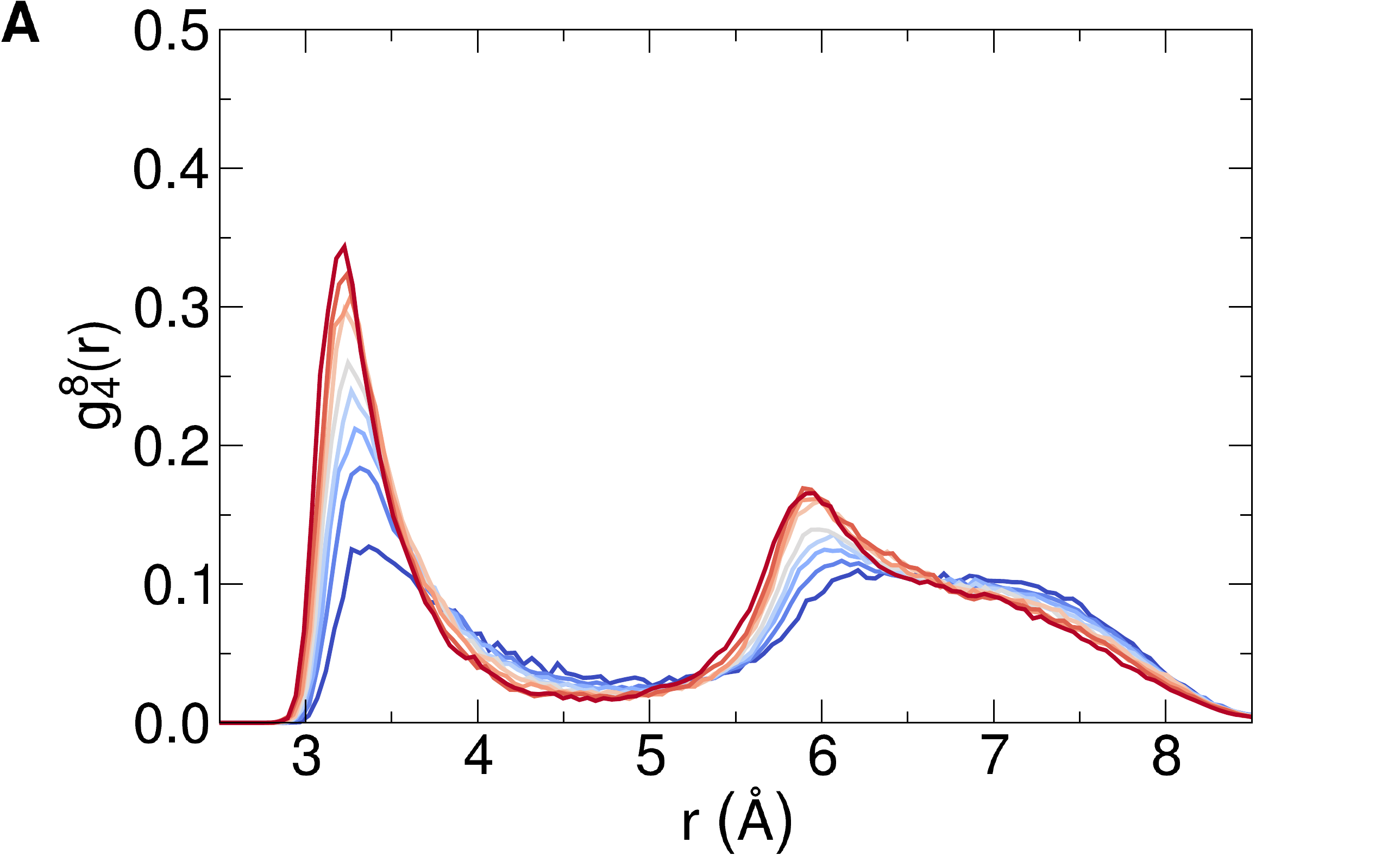}
	\includegraphics[width=8cm]{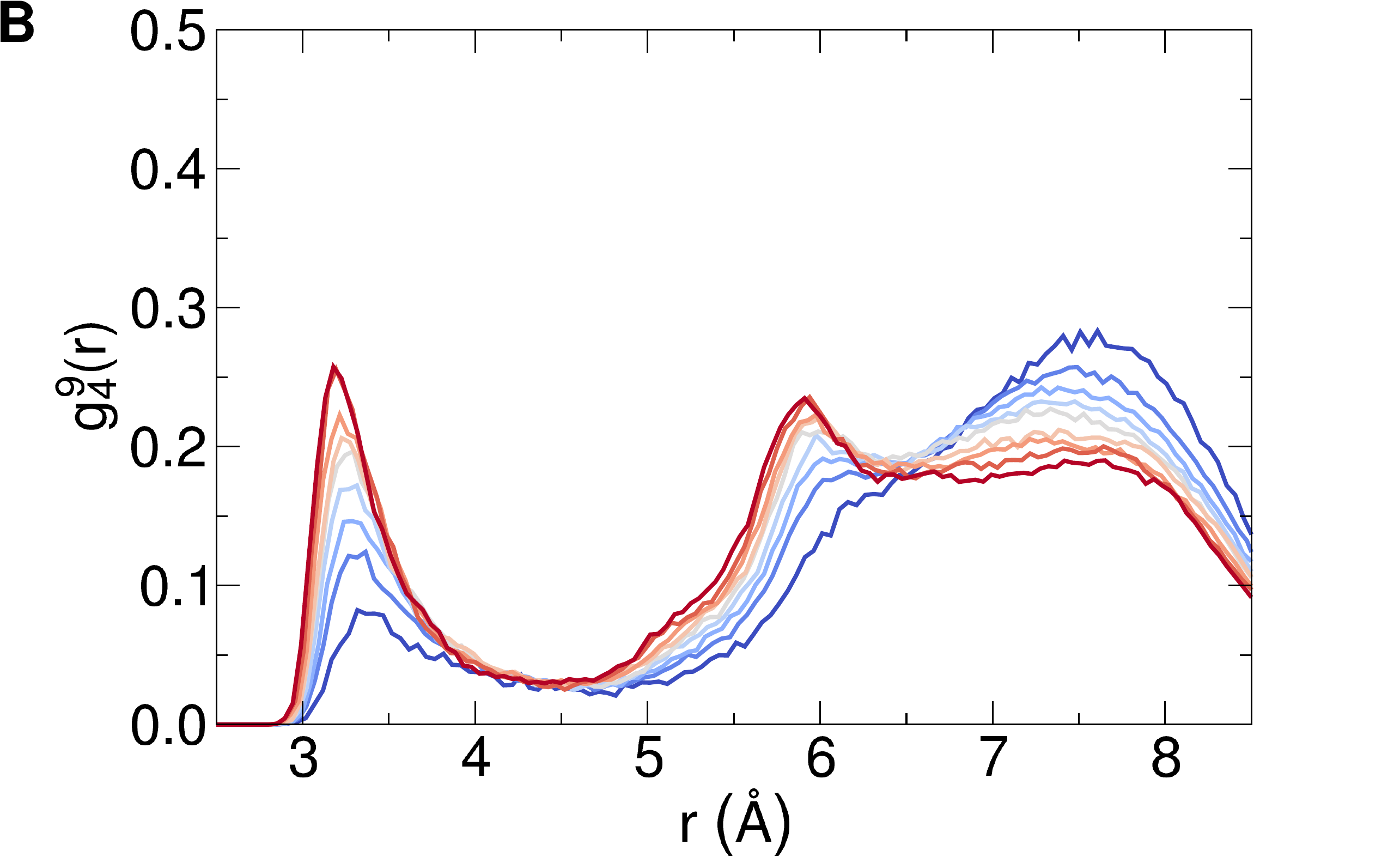}
	\vskip 6pt
	\includegraphics[width=8cm]{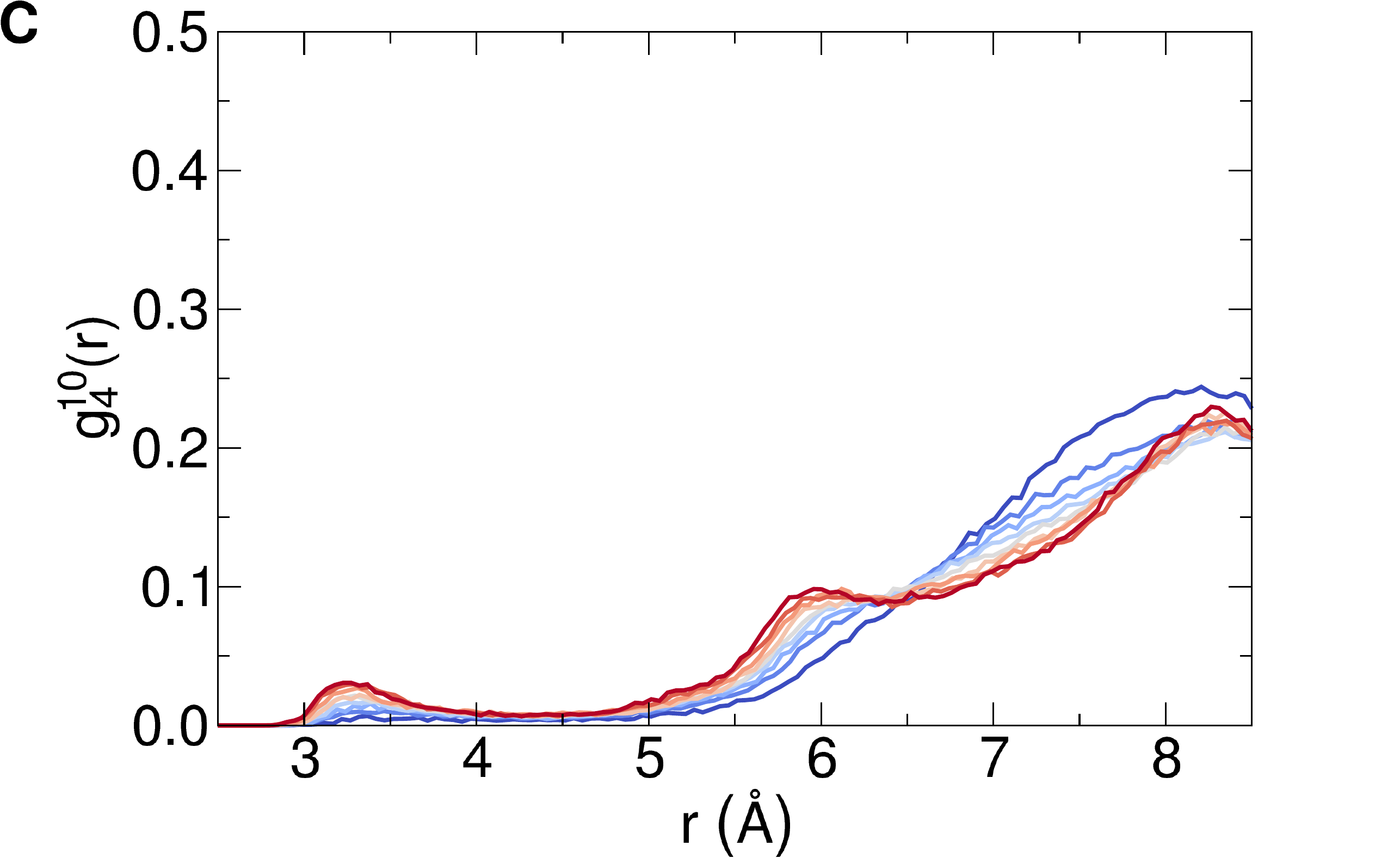}
	\caption{Contributions to the \( g(r) \) arising from pairs of
	molecules at chemical distance \( D=4 \) in rings of length
	\( L \) ranging from (a) \( L=8 \) to (c) \( L=10 \).
	Pressure increases from blue to red.}
	\label{fig:grringsD4}
\end{figure}

The pressure dependence of each \( g_D^L \) along the isotherm is shown in Figs.~\ref{fig:grringsD2}--\ref{fig:grringsD4}.
The behavior of \( g_2 \) and \( g_4 \) (and their components \( g_2^L \) and \( g_4^L \)) is extensively discussed in the main text. Besides the increase in the interstitial peak, \( g_3 \) does not show clear relevant features in the evolution from high to very-high densities.
As seen in Fig.~\ref{fig:grringsD3}, the (small) increase in the \( g_3 \) contribution at \( r\approx\SI{6}{\angstrom} \) can be ascribed fundamentally to the \( g_3^8 \) component, where a peak continuously shifting from \( \approx\SI{6.5}{\angstrom} \) to \( \approx\SI{6}{\angstrom} \) is observed, without evidence of significant transformations in the underlying network structure.

\clearpage

%